\documentclass[useAMS,usenatbib]{mnras}
\usepackage{graphicx}
\usepackage{color}

\newcommand{\ml}{$\rm M/L$}
\newcommand{\mlr}{$\rm M_\star/L_r$}

\newcommand{\simlt}{\lower.5ex\hbox{$\; \buildrel < \over \sim \;$}}
\newcommand{\simgt}{\lower.5ex\hbox{$\; \buildrel > \over \sim \;$}}

\newcommand{\mgf}{$\rm Mg4780$}
\newcommand{\atio}{$\rm aTiO$}
\newcommand{\tioi}{$\rm TiO1$}
\newcommand{\tioiio}{$\rm TiO2_{SDSS}$}
\newcommand{\tioii}{$\rm TiO2$}

\newcommand{\fem}{$\rm \langle Fe\rangle$}

\newcommand{\nai}{$\rm NaD$}
\newcommand{\nad}{$\rm NaD$}
\newcommand{\naii}{$\rm NaI8190$}

\newcommand{\cat}{$\rm CaT$}
\newcommand{\cai}{$\rm Ca1$}
\newcommand{\caii}{$\rm Ca2$}
\newcommand{\cnii}{$\rm CN2$}
\newcommand{\caiii}{$\rm Ca3$}
\newcommand{\cfs}{$\rm C4668$}
\newcommand{\mgi}{$\rm Mg1$}
\newcommand{\mgii}{$\rm Mg2$}
\newcommand{\mgfep}{$\rm [MgFe]'$}
\newcommand{\mgfe}{$\rm [Mg/Fe]$}
\newcommand{\cafe}{$\rm [Ca/Fe]$}
\newcommand{\nafe}{$\rm [Na/Fe]$}
\newcommand{\tife}{$\rm [Ti/Fe]$}

\newcommand{\cfe}{$\rm [C/Fe]$}
\newcommand{\ofe}{$\rm [O/Fe]$}
\newcommand{\nfe}{$\rm [N/Fe]$}
\newcommand{\sife}{$\rm [Si/Fe]$}

\newcommand{\feff}{$\rm Fe4531$}
\newcommand{\hbo}{$\rm H\beta_o$}
\newcommand{\hb}{$\rm H\beta$}

\newcommand{\mgb}{$\rm Mgb5177$}
\newcommand{\hgf}{$\rm H\gamma_F$}
\newcommand{\caf}{$\rm Ca4227$}
\newcommand{\cafr}{$\rm Ca4227r$}
\newcommand{\kms}{\,km\,s$^{-1}$}
\newcommand{\afe}{$[\rm \alpha/{\rm Fe}]$}

\newcommand{\gammab}{$\rm \Gamma_b$}
\newcommand{\xfe}{$\rm [X/Fe]$}
\newcommand{\zh}{$\rm [Z/H]$}

\newcommand{\alfa}{$\alpha$}

\newcommand{\VROT}{$\rm V_{rot}$}
\newcommand{\SIG}{$\rm sigma$}
\newcommand{\HT}{$\rm H_3$}
\newcommand{\HF}{$\rm H_4$}

\usepackage[T1]{fontenc}
\usepackage{ae,aecompl}

\usepackage{graphicx}	% Including figure files
\usepackage{amsmath}	% Advanced maths commands
\usepackage{amssymb}	% Extra maths symbols

\title[IMF of the M31 bulge]{Mild radial variations of the stellar IMF in the bulge of M31.}

\author[F. La Barbera et al.]
{F. La Barbera$^{1}$\thanks{E-mail:  francesco.labarbera@inaf.it (FLB)},
A., Vazdekis$^{2,3}$, 
I. Ferreras$^{2,3,4}$, 
A. Pasquali$^{5}$\\
$^{1}$INAF-Osservatorio Astronomico di Capodimonte, sal. Moiariello
16, Napoli, 80131, Italy\\
$^{2}$Instituto de Astrof\'\i sica de Canarias, Calle V\'\i a L\'actea s/n, E-38205
  La Laguna, Tenerife, Spain\\
$^{3}$Departamento de Astrof\'\i sica, Universidad de La Laguna (ULL), E-38206  La Laguna, Tenerife, Spain\\
$^{4}$Department of Physics and Astronomy,  University College London,  Gower Street,  London WC1E 6BT, UK\\
$^{5}$Astronomisches Rechen-Institut, Zentrum f\"ur Astronomie, Universit\"at Heidelberg, M\"onchhofstr. 12-14, D-69120 Heidelberg, Germany\\
  }

\date{}

\pubyear{2021}

\begin{document}
\label{firstpage}
\pagerange{\pageref{firstpage}--\pageref{lastpage}}
\maketitle

\begin{abstract}
Using new, homogeneous, long-slit spectroscopy in the wavelength range from $\sim 0.35$ to $\sim 1 \, \mu$m,
we study radial gradients of optical and near-infrared (NIR) IMF-sensitive features along the major axis of the bulge of M31, out to a galactocentric distance of $\sim 200$~arcsec ($\sim 800$~pc).  Based on state-of-the-art stellar population synthesis models with varying Na abundance ratio, we fit a number of spectral indices, from different chemical species (including TiO's, Ca, and Na indices), to constrain the low-mass ($\lesssim 0.5$~$M_\odot$) end slope (i.e. the fraction of low-mass stars) of the stellar IMF, as a function of galactocentric distance.
Outside a radial distance of $\sim 10$'', we infer an IMF similar to a Milky-Way-like distribution, while at small galactocentric distances, an IMF radial gradient is detected, with a mildly bottom-heavy IMF in the few inner arcsec. 
We are able to fit Na features (both NaD and \naii ), without requiring extremely high Na abundance ratios. \nafe\ is $\sim 0.4$~dex for most of the bulge, rising up to $\sim 0.6$~dex in the innermost radial bins.
Our results imply an overall, luminosity-weighted, IMF and mass-to-light ratio for the M31 bulge, consistent with those for a Milky-Way-like distribution, in contrast to results obtained, in general, for most massive early-type galaxies. 
\end{abstract}


%%%%%%%%%%%%%%%%%%%%%%%%%%%%%%%%%%%%%%%%%%%%%%%%%%

%%%%%%%%%%%%%%%%% BODY OF PAPER %%%%%%%%%%%%%%%%%%

\section{Introduction}
The  Initial  Mass   Function  (IMF),   i.e.   the distribution of  the masses  of stars, at  birth, in a stellar system, is a key ingredient  of astrophysics, as it sets  the  overall  mass-scale   of  galactic  systems,  controls  the intensity of  the stellar feedback  processes and drives  the chemical enrichment abundance patterns. While the IMF has long been assumed to be ``universal'', and the same as in the solar neighborhood, a number of observational studies have found evidence of IMF variations for both star-forming and  quiescent systems \citep[see][for a recent review]{AH:18}. Understanding the origin of these variations  is one of the main challenges that nowadays the astronomical community is facing on.

In  early-type  galaxies  (hereafter  ETGs),
\citet{Capp:12, Capp:13} have found a systematic increase in the stellar $\rm M/L$ with galaxy mass, based on detailed dynamical models of the galaxies' kinematics, with a Kroupa IMF normalization at low  velocity dispersion ($\sigma\sim 80$\,\kms) transitioning to
a Salpeter, i.e. bottom-heavier than the Milky-Way, one at $\sigma\sim  260$\,\kms. Independent dynamical
studies    have   achieved similar    conclusions   \citep[see,
  e.g.,][]{TMJ11,  WCT:12,  Capp:12, Dutton:12,Tortora:13}.
Strong gravitational lensing over  galaxy scales can also be exploited
to  constrain the stellar  $\rm M/L$. While strong  lensing studies  of  low-mass  spheroids 
excluded a Salpeter IMF in favour of a Milky-Way-like distribution~\citep{FSW:05,FSB:08,ECross:10}, other works have detected a systematic  variations towards higher  stellar $\rm M/L$  with increasing  mass~\citep{Auger:10, Treu:10,
  Barnabe:11}, with some exceptions~\citep{SmithLucey2013, SLC:2015, Leier:2016}.  In general,  most lensing  and dynamical studies have consistently pointed to a
scenario where the IMF is either bottom- or  top-heavier  (both implying { a higher} $\rm M/L$) than
Kroupa/Chabrier, in  massive ETGs.  Indeed, the analysis of spectral features sensitive to the presence of low-mass stars, such as the Na\,I doublet feature at
$\lambda\lambda8183,8195$\,\AA\ \citep[hereafter
  NaI8200]{FaberFrench:80, SchiavonFeH:97},  have shown that the above M/L variations are likely due to a bottom-heavy IMF, i.e. an enhanced relative contribution of dwarf
versus giant stars, in most massive ETGs (see, e.g., \citealt{Cenarro:2003,vdC:10,Ferr:13,LB:13, Spiniello:2014}; but see~\citealt{Alton:2017, Alton:2018}). Unfortunately, all these studies lacked spatial resolution within galaxies, constraining only the ``integrated'' (light-weighted) IMF, on apertures encompassing most of the galaxy total light. Since, overall, both velocity-dispersion, elemental abundance ratios, age, and stellar metallicity increase as a function of galaxy mass in ETGs (see, e.g., ~\citealt{Gallazzi:20}, and references therein), studies of integrated galaxy properties are not able to fully disentangle the effect of different possible drivers of IMF variations~\citep{LB:15}.  

Therefore, subsequent studies have focused on radial gradients of the IMF in ETGs
\citep{NMN:15a,NMN:15b,LB:16,LB:17,Ziel:17,vanDokkum:2017,Parikh:2018,Sarzi:2018, Dominguez:2019, LB:19}, finding that a bottom heavy IMF is only present in the central regions of the most massive galaxies. This result seems to be also consistent with a two-phase scenario of galaxy formation~\citep{Oser:10}. Although radially resolved studies have allowed us to enlarge significantly the parameter space where IMF variations can be constrained, the problem is far from being sorted out. 
Based on spatially-resolved measurements of the TiO2 IMF-sensitive features for ETGs in the CALIFA survey,
\citet{NMN:15b} found that metallicity, rather than {\sl local} velocity
dispersion, might be the primary {\it local} driver of IMF variations.  
However, \citet[hereafter LB19]{LB:19} found that in most massive ETGs, mostly brightest cluster galaxies, regions with high metallicity do not necessarily exhibit a bottom-heavy IMF, while \citet{NMN:19} showed that in a disk-dominated galaxy, IMF variations do not mimic metallicity changes. 

The bulges of spiral galaxies might allow us to further expand the parameters' space where IMF variations can be measured through spectral features. At least some bulges have high metallicity and old ages in their central regions, similar to those of massive ETGs (see, e.g., fig.~3 of~\citealt{TD:06}), but lower velocity dispersion (a proxy for galaxy mass). Because of { their lower mass}, bulges set tighter upper limits to possible IMF variations, and thus offer a good benchmark to pinpoint possible systematics in the analysis of IMF-sensitive spectral features. The bulge of M31, due to its proximity, gives a unique opportunity to obtain high S/N spectroscopy, at low observational cost, and apply the same methodology used so far to constrain the stellar IMF of ETGs. Indeed, the stellar populations in the bulge are old, enhanced in alpha elements, and have super-solar metallicity in the center, similar to massive ETGs~\citep{Saglia:2010, Saglia:2018}. On the contrary, { the bulge's} velocity dispersion is low ($\sim 150$~\kms). 

The stellar IMF of the M31 bulge has been the subject of an intense debate over the years, since the 1970s. \citet{SpinTa:71} measured a very strong \naii\ feature, explaining it through a dwarf-dominated IMF, with a high mass-to-light ratio ($\rm M/L=44$). A similar result was obtained by ~\citet{Oconnell1976}, but with lower M/L (between 2 and 15). On the contrary, based on NIR spectroscopy targeting the FeH Wing-Ford band~\citep{SchiavonFeH:97} and the K-band CO absorption, \citet{Whitford:77} and~\citet{Cohen:1978}
concluded that the bulge IMF is similar to that of the solar-neighborhood. The debate has continued over the years (see, e.g., ~\citealt{Carter:1986, DelisleHardy:1992}), but it is only with the advent of modern instrumentation and improved stellar population models that 
we are now { in the position} to make a significant leap forward.

Indeed,  recent studies have shown that the bulge of M31 has a much lighter IMF than what suggested in the 1970s.
~\citet[hereafter CvD12b]{CvD12b} analyzed spectra  for  the innermost few arcsec region of M31. Performing spectral fitting in the ranges $0.4 < \lambda < 0.55$~$\mu$m and $0.8< \lambda < 1.02$~$\mu$m with stellar population models including the effect of non-solar abundance ratios, CvD12b found evidence for an IMF normalization between Kroupa and Salpeter ($\rm M/L_r \sim 6.2$). 
\citet[hereafter Z15]{Z:15} analyzed NIR spectral features (NaI8200, CaT, and FeH) at six radial positions in the bulge of M31 (out to a radial distance of 700''), performing a qualitative comparison of observed line-strengths to SSP model predictions from~\citet[hereafter CvD12a]{CvD12a}. Z15 concluded that M31 is well described by a Chabrier IMF at all radial positions, hence favouring a lighter IMF in the center, with respect to CvD12b. Both CvD12b and Z15 concluded that the strong NaI8200 absorption in the center of M31 is likely due to Na over-abundance, rather than IMF (as originally proposed by~\citealt{SpinTa:71}), with \nafe\ as high as $\sim 1$~dex. The NaI8200 radial gradient would then be explained as a strong \nafe\ radial gradient (see Z15). However, an extremely high value of \nafe\ is not without problems. In the center of the most massive ETGs, LB19 found \nafe\ abundances not  higher than $\sim 0.6$--$0.7$~dex, even in the most metal-rich regions. \citet{Bensby:17}, analyzing stars in the Galactic bulge, have found, for individual stars, \nafe\ abundances  below $\sim 0.3$~dex, even at the highest metallicity probed ($\rm [Fe/H] \sim 0.5$~dex), far below the above estimate for the center of M31. 
We point out that \nafe\ has been measured indirectly by CvD12b (i.e. based on model predictions for the effect of \nafe\ on other spectral features), as they lacked the region around NaD, the most sensitive feature to \nafe\ in optical galaxies' spectra. Instead, Z15 used \nad\ and \naii\ measurements from different sources, with \nad\ from ~\citet{Davidge:1991, Davidge:1997}.
Therefore, the question remains open, about the origin of Na absorption in the center of M31. 
     
The present work presents the most detailed, homogeneous, study of IMF-sensitive spectral features along the major axis of the M31 bulge, from the optical throughout NIR spectral range. To this effect, we have acquired new, dedicated, high-S/N, spectroscopy along the major axis of the bulge, with the OSIRIS  (Optical System for Imaging and low-Intermediate-Resolution Integrated Spectroscopy) spectrograph at the GRANTECAN (GTC) telescope, ensuring a continuous, homogeneous, spectral coverage from $\lambda \sim 0.35 \, \mu$m, to $\lambda \sim 1  \, \mu$m. Contrary to previous works, we target {\it all} IMF-sensitive features up to $\lambda \sim 1 \mu$m, including Mg, Fe, TiO, Na (both NaD and \naii), and Ca absorptions, and analyze them based on state-of-the-art stellar population models that take the {\it coupled effect} of IMF and Na abundance explicitly into account, applying the same approach that we have used so far to infer the stellar IMF of massive ETGs (see,  e.g., ~LB19 and references therein).
     
The layout of the paper is as follows.  In Sec.~\ref{sec:specdata}, we
describe the new OSIRIS data for the M31 bulge, the kinematics extracted from the new spectroscopy, as well as the radially binned spectra used in our analysis.  
Sec.~\ref{sec:spmodels} describes the stellar population models used to analyze the spectra. The analysis is presented in Sec.~\ref{sec:analysis}, including the definition of spectral indices used in this work (Sec.~\ref{sec:defindices}), the methods used  to constrain IMF and other stellar population properties (Sec.~\ref{sec:agez} and~\ref{sec:IMFfitting}), as well as the comparison of model and best-fitting spectral indices (Sec.~\ref{sec:bestfit}). Sec.~\ref{sec:results} shows the main results of the present work, and in particular,
the IMF and \nafe\ radial profiles for the bulge of M31 (Sec.~\ref{subsec:IMFgradient} and~\ref{subsec:NaFegradients}, respectively).
We discuss the results in Sec.~\ref{sec:Discussion}. Summary and conclusions follow in Sec.~\ref{sec:Conclusions}.
Throughout the present work, we adopt a distance of 785~kpc from the MW to M31~\citep{McConnachie:2005}, implying a conversion scale of $\rm \sim 3.8$~pc/arcsec.

\section{New UVRI spectroscopy for the M31 bulge}
\label{sec:specdata}

\subsection{Observations}
\label{subsec:obs}
We obtained new long-slit spectroscopy for the bulge of M31 on August and September 2017, using the OSIRIS instrument at the Nasmyth-B focus of the Gran Telescopio CANARIAS (GTC), at Roque de los Muchachos Observatory. OSIRIS is equipped  with a mosaic of two 2k$\times$4k red-optimized CCDs, separated by a narrow gap of 9.4''. Observations were carried out with the U-, V-, R-, and I- R2500 grisms, using a 7.4'-long slit of width 0.4'', resulting { in a} spectral coverage from $3460 $  to $10150$~\AA\ with a uniform spectral resolution of $\sim 38$~\kms\ (FWHM; as measured from arc lamps and sky emission lines). The M31 pointings were centered at RA=00:42:44.57 and DEC=41:16:05.7, with the slit aligned along the bulge major axis, at PA=48~deg~(see~\citealt[hereafter S10]{Saglia:2010}). The target was centered on CCD\#2, 55'' apart from the gap, allowing us to observe the bulge up to a distance of $\sim 180''$ ($\sim 260''$) along the East (West) direction. We adopted a 2x (1x) binning along the spatial (dispersion) direction, resulting { in a} spatial scale of $0.254''$/pixel, and a dispersion of $\sim 0.3$, $\sim 0.4$,  $\sim 0.5$, and $\sim 0.7$~\AA\ per pixel for the UVRI grisms, respectively.  For each grism, we performed two observations of the bulge, each followed by a pointing on a blank sky region, centered at RA=00:44:27.20 and DEC=40:49:54.8. For the VRI grisms, each { observation} consisted of three dithered exposures of 150sec each, with a dithering pattern of $\sim$15{''},  resulting { in a} total on-target exposure time of 900sec. For the U grism, we performed five exposures, giving a total integration time of 1500sec. 
The average seeing was 1, 1.2, 1.1, and $1''$ (FWHM) for the UVRI grisms, respectively.

\subsection{Data reduction}
\label{subsec:reduction}
The data were reduced using dedicated FORTRAN and IRAF scripts written by the authors. For each CCD, all frames were bias subtracted, trimmed, and flat-fielded using twilight sky frames. For the I grism, wavelength calibration was performed using sky lines only, while for the V and R grisms, we combined sky lines with HgAr+Ne+Xe arc lamp lines. For the U grism, wavelength calibration was performed with HgAr+Xe calibration lamp lines. For the R and I grisms, we estimated the accuracy of the wavelength calibration by measuring the position of sky lines in the calibrated frames, finding an rms (across the spatial direction and the spectral range of each grism) of $\sim 4$~\kms .
Flux calibration was performed using data for a spectrophotometric standard star  observed with the same instrumental setup as for the science exposures. For each grism, the flux standard was observed right before the observations of M31.
For the R and I grisms, a standard star was also observed at different positions on both OSIRIS CCDs, in order to map spatial variations of the response function across the entire field of view. We found variations of up to 10$\%$ at the edges of each spectral range, among the response functions of CCD\#2 and CCD\#1. These response functions were interpolated across the spatial direction and used to further improve the flux calibration.

Sky subtraction was performed by correcting each M31 observation with the corresponding sky frame. Sky lines were rescaled to match those in the M31 frame by using the software {\sc SKYCORR}~\citep{Noll:2014}. To this aim, each M31 frame was binned along the spatial direction, and a one-dimensional spectrum was extracted from the M31 and the corresponding sky frames. For each bin, we ran SKYCORR to derive the corresponding sky-lines rescaling factors. Interpolating the scaling factors across the spatial direction provided us with a rescaled 2D sky spectrum, that was subtracted off from the given M31 frame. Particular care was taken in the subtraction of the NaD sky lines at $\lambda\lambda \sim 5890, 5896$~\AA. The (prominent) NaD stellar absorption in M31 is potentially contaminated by dust absorption due to the ISM in M31 and our Galaxy { (see App.~\ref{app:ism} for details)}. Since the ISM absorption from our Galaxy overlaps, in wavelength, to the NaD sky lines, rescaling the NaD in the sky spectra with those in the M31 frames may lead to underestimate the true intensity of the NaD sky lines. Since we found that the intensity of the NaD sky lines was stable among different observations, we decided to not apply any rescaling. Instead, we modeled the NaD lines in each sky frame with a combination of Gaussian functions,  applied a shift to match the position of the lines in the M31 frame (as a function of the spatial direction along the slit), and subtracted them off. Before sky subtraction, all frames were corrected for bad pixels and cosmic rays,  by performing a linear interpolation of affected pixels along the slit direction. 

For the R- and I- grism, each sky-subtracted frame was corrected for telluric lines, using the software {\sc MOLECFIT}~\citep{Smette:2015, Kausch:2015}. {\sc MOLECFIT} produces a theoretical transmission model by fitting selected regions, with prominent telluric lines, in the object spectrum itself. We ran  {\sc MOLECFIT} on a spectrum extracted in the innermost region of M31, verifying that results were unchanged when considering spectra extracted at different positions along the bulge.

For each observation, the sky-corrected M31 frames of both CCDs were joined together, rectified (i.e. corrected for spatial variations of the photometric center of the galaxy along the dispersion direction) and combined using the IRAF task {\sc IMCOMBINE}. At each reduction step, and for each science frame, a variance map was produced and updated during the reduction process to account for different sources of { uncertainty, including that on sky subtraction\footnote{ To estimate the uncertainty on the correction of NaD sky lines, we repeated the subtraction using different sky frames, and included the corresponding error budget { in the} variance maps of sky-subtracted frames.}}. Variance maps were finally combined in the same way as for the science data. 

\subsection{Radially binned spectra}
\label{subsec:binning}

For each grism, we extracted radially binned spectra along both sides of the slit. The central bin was set to be 
1.5$''$-wide around the photometric center of the galaxy, 
while for the other bins we adopted a minimum width of $0.75''$ ($\sim 3$ pixels). When required, the bin width was adaptively increased outwards, to ensure a minimum signal-to-noise ratio, $\rm S/N_{min}$. The S/N was computed in the U-grism spectral range, from 3900 to 4200~\AA\ (corresponding to the CaH+K absorption lines), as the S/N was found to be higher at redder wavelengths. Since we did not perform absolute flux calibration, the binned spectra from different grisms were just joined by applying suitable multiplicative factors, derived from overlapping spectral regions.

We extracted two sets of binned spectra. First, we set $\rm S/N_{min}=25$~\footnote{We also repeated the analysis with $\rm S/N_{min}=15$, finding consistent results to those for $\rm S/N_{min}=25$.}
 to derive the kinematics of the bulge, i.e. radial profiles of rotation velocity \VROT , velocity dispersion, \SIG , and higher moments of the line-of-sight (LOS) velocity distribution, \HT\ and \HF, respectively. As detailed in App.~\ref{app:kin}, the kinematics was extracted with the software {\sc pPXF}~\citep{Cap:2004, Capp:17}, performing spectral fitting on different spectral regions, and combining results into final radial profiles. In App.~\ref{app:kin}, we show that the new kinematics is in good agreement with that obtained by previous studies.

At each position along the spatial direction, the rotation velocity profile was interpolated and used to correct the two-dimensional spectra of M31 to the restframe. To perform the stellar population analysis (see below), a second set of radially binned spectra was then extracted with $\rm S/N_{min}=70$~\footnote{We repeated the analysis with $\rm S/N_{min}=90$, finding consistent results { with those for $ \rm S/N_{min}=$70}. Note that adopting $\rm S/N_{min}=70$ ensured a S/N ratio larger than $\sim100$~\AA$^{-1}$ for all the IMF sensitive features analyzed in this work, allowing both IMF and abundance ratios to be properly constrained (see, e.g.,~\citealt[hereafter LB13]{LB:13}).}, in order to perform the stellar population analysis, as detailed below. 

\begin{figure*}
 \begin{center}
\leavevmode
\includegraphics[width=17cm]{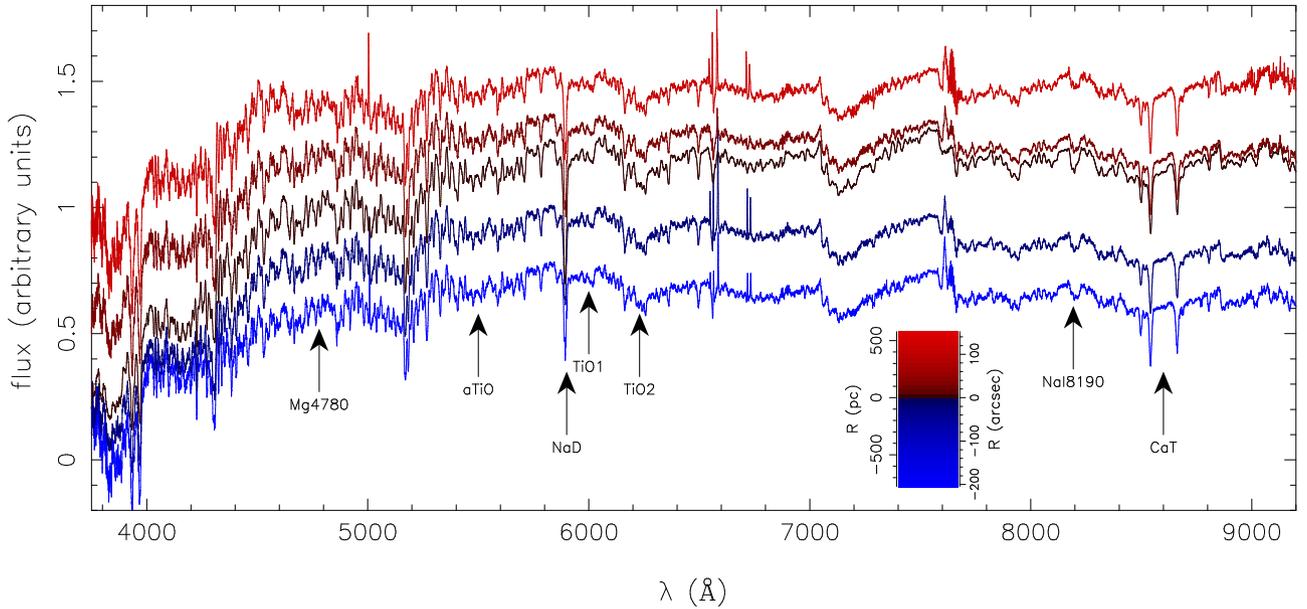}
 \end{center}
    \caption{The figure plots, as an example, some of the M31 radially binned spectra used for the stellar population analysis (with $\rm S/N_{min}=70$; see the text). The distance to the center along the slit, $\rm R$,  is encoded with different colours, from blue (North-East side of the slit) through red (South-West), as shown by the inset coloured bar. From bottom to top, the plot shows the M31 spectra at distances of $R \sim $ -205, -27, 0 (i.e. the central spectrum), +28, and +152~arcsec ($\sim -787$, -103, 0, 106   and 578~pc, respectively) from the galaxy center. The bluest and reddest spectra correspond to the largest galactocentric distances probed with our stellar population analysis. The spectra have been normalized to one in the spectral range from 4800 to 5400~\AA, and arbitrarily shifted for displaying purposes. The IMF-sensitivy features analyzed in the present work are marked with black arrows.
    }
    \label{fig:M31spec}
\end{figure*}

Fig.~\ref{fig:M31spec} plots, as an example, some of the $\rm S/N_{min}=70$ radially binned spectra of M31, showing the excellent quality of the new OSIRIS data.

\section{Stellar population models}
\label{sec:spmodels}

To analyze the spectra of M31, we rely on Na--EMILES and $\alpha$--enhanced MILES stellar population models.

The Na--EMILES models are  a dedicated version  of the EMILES models, covering a range of \nafe\ abundance ratios (see LB19 and references therein). EMILES  simple stellar population (SSP) models cover the spectral range from $0.35$ to $5 \, \mu$m, based on different empirical stellar libraries, namely MILES in the optical range \citep{MILESI}, up to $\lambda  \! \sim \! 7410$~\AA , Indo-US~\citep{Valdes04} and 
CaT~\citep{CATI} out to $\lambda \! \sim \! 8950$~\AA\ (\citealt{Vazdekis:12}), and the IRTF stellar library~\citep{IRTFI,IRTFII} at redder wavelengths (see \citealt{RV:16} for details). 
The spectral resolution is kept constant with wavelength (at FWHM=2.5~\AA) for all libraries, except for IRTF, having a constant $\sigma$=60~\kms\ (see figure~8 of~\citealt{Vazdekis:2016}). Note that the above spectral libraries follow the abundance pattern of the Milky-Way, (i.e. they are approximately scaled-solar at solar metallicity, and significantly $\alpha$--enhanced at metallicity below \zh\ about $-0.3$~dex). Hence, EMILES SSPs should be considered as ``base'' (not scaled-solar) models.
The Na--EMILES models are computed for two sets of scaled-solar theoretical isochrones, namely
the ones of \citet{Padova00}
(Padova00) and those of \citet{Pietrinferni04} (BaSTI), the latter having cooler temperatures  at the low-mass end~(see \citealt{Vazdekis:15}, and references therein, for details).  To cover a range in \nafe , we apply theoretical differential corrections for \nafe\ overabundance to each individual stellar spectrum in the
empirical libraries, and construct the Na--EMILES SSPs based 
on scaled-solar isochrones (see~\citealt[hereafter LB17]{LB:17}, and LB19). { Note} that this approach differs from that used to construct $\alpha$--enhanced MILES models (hereafter $\alpha$--MILES; see V15), where 
corrections for non-solar \afe\ abundance ratios were applied directly to the 
model SSPs, rather than to individual stars. Also, the \alfa--enhanced
models only cover the optical (MILES) spectral range and are computed for BaSTI isochrones only, while Na--enhanced models are computed over the optical plus NIR spectral range for both Padova00 and BaSTI isochrones. 

Models are computed for different values of age, from $1.0$ to $\sim 14$\,Gyr, and total metallicity, \zh . For BaSTI SSPs (either Na--EMILES or $\alpha$--enhanced MILES), five values of total metallicity are considered, i.e. $\rm [Z/H]=\{$ $-0.66$, $-0.35$, $-0.25$, $0.06$, $+0.26\}$~\footnote{Note that BaSTI models are also computed for \zh$=0.4$. However, given the lower quality of these models they are not used in the present analysis.}, while for Padova00, we rely on models having $\rm [Z/H]=\{$ $-0.71$, $-0.4$, $0$, $+0.22\}$, respectively.  
The Na--EMILES models are computed for \nafe$=$\{$0, 0.3, 0.6$, $0.9$\}~dex, while $\alpha$--enhanced models are given for \afe$=0$ (solar scale), and \afe$=+0.4$~dex.

Na--EMILES and $\alpha$--enhanced MILES models are computed for different IMF shapes, and in particular for two power-law distributions, i.e { a single power-law and a low-mass tapered single power-law IMF, also referred to as ``unimodal'' and ``bimodal'' distributions, respectively} (see \citealt{CATIV}, V15, and \citealt{Vazdekis1996}). 
The { single power-law and low-mass tapered} IMFs are defined by their logarithmic slopes, $\Gamma$ and $\Gamma_b$,
respectively. The lower and upper mass-cutoffs of the IMF are set to $0.1$ and $100$\,M$_\odot$, respectively. 
{ We note that} the difference between a { single power-law} and a { low-mass tapered IMF is that the latter} is smoothly tapered towards { masses below} $\sim 0.5 \, M_\odot$; hence, varying { the slope $\Gamma_b$ changes the ratio of dwarf-to-giant stars} in the IMF through its overall normalization. While this approach is different with respect to a change of the low-mass slope (e.g.~\citealt{CvD12b}), { we emphasize that a low-mass tapered} parametrization is
suitable for our purposes, as most IMF-sensitive features { are only sensitive to} the dwarf-to-giant ratio in the IMF (e.g.~LB13, ~\citealt[hereafter LB16]{LB:16}). { Moreover, a low-mass tapered IMF has been shown to provide mass-to-light ratios more consistent with dynamical constraints~\citep{Lyubenova}, and to be able to describe both optical and NIR IMF-sensitive features in galaxy spectra} (see LB16). { Therefore, in the present work, we consider only models with a low-mass tapered distribution, that do also allow us to perform a direct comparison to previous works for massive galaxies (see Sec.~\ref{sec:Discussion}).} 
The Na--EMILES and $\alpha$--MILES models are computed for the following { low-mass tapered} IMF slopes, \gammab= \{$0.3, 0.5, 0.8, 1.0, 1.3, 1.5, 1.8, 2.0, 2.3, 2.5, 2.8, 3.0, 3.3, 3.5$\}. For $\Gamma_b=1.3$, the {low-mass tapered} IMF closely approximates the~\citet{Kroupa01}
Universal IMF.

\section{Stellar population analysis}
\label{sec:analysis}

\subsection{Definition of spectral indices}
\label{sec:defindices}
The wide spectral range provided by the OSIRIS UVRI grisms allows us to probe a wide set of optical and NIR spectral features. Following the same approach as in our previous works (e.g. LB13, LB17, and LB19), we select a set of optical and NIR spectral indices, and constrain the stellar population properties of M31 by comparing their observed line-strengths to model predictions. We consider the age-sensitive Balmer indices, \hbo\ and \hgf, the total metallicity indicator \mgfep , as well as the IMF-sensitive features \tioi, \tioiio, \atio, and \mgf, the two Na indices, \nai\ and \naii, and the Calcium triplet lines, \cai, \caii, and \caiii, with the combined \cat=$\rm Ca1 \! + \! Ca2 \! + \! Ca3$~\citep{CATI}. We also include other indices in the analysis, such as \cafr, \cfs, \cnii, \feff, Fe5015, \mgb, \mgi, \mgii, which are mostly sensitive to abundance ratios (see table~1 of LB15), and for which we adopt the same central passband and pseudo-continua definitions as in the Lick system~\citep{Trager98}.  The index definitions for \tioi, \hgf, and \nai\ are also the same as in \citet{Trager98}, while \hbo\ is the optimized \hb\ index defined by \citet{CV09}. The \tioiio\ is defined as in LB13, being a modified version of \tioii\ from \citet{Trager98}. 
The total-metallicity indicator \mgfep$\rm =[ Mgb5177 \cdot (0.72 Fe5270+0.28 Fe5335) ]^{1/2}$ is a combined Mgb and Fe index, defined by~\citet{TMB:03} to be insensitive to \mgfe\ abundance ratio (see also V15). Finally, the \mgf\ is from~\citet{Serven:2005}, \atio\ from~\citet{Spiniello:2014}, while \naii\ is defined as in~CvD12a, with some modifications as described in LB17.

{ We note} that (i) the spectral indices \hbo, \hgf, \mgb, and Fe5015, are corrected for contamination from emission lines, as detailed in App.~\ref{app:ecorr}; (ii) for each radial bin, when fitting observed and model line-strengths (see Sec.~\ref{sec:agez} and~\ref{sec:IMFfitting} below),  models are first smoothed to match the sigma~\footnote{We refer to sigma here to indicate the root square of the sum in quadrature of the velocity dispersion of a given spectrum and the instrumental resolution.} of the given 
bin~\footnote{This approach maximizes the information we can extract from the data, as it does not require any smoothing of the spectra to bring all of them to the same sigma. However, in all Figures of the present paper, to compare observed and model line-strengths among different radial bins, we correct all line-strengths to the same sigma of $\sigma_0=$150~\kms. The correction is given as the index variation between a reference SSP model smoothed to $\sigma_0$ and to the sigma of a given bin. As a reference model we use, for each radial bin, the best-fitting model from method A (see Sec.~\ref{sec:IMFfitting}).}; (iii) NaD was corrected for dust contamination, as detailed in App.~\ref{app:ism}.
 
\subsection{Constraining age, metallicity, and \mgfe}
\label{sec:agez}

As a first step in the analysis, we estimate the age, metallicity, and \mgfe\ abundance ratio. We adopt different methods:
\begin{description}
 \item[ i. ] We use 1SSP model predictions from EMILES base models with  Padova00 isochrones (hereafter, EMILES iP), for a Kroupa-like IMF (i.e. a {low-mass tapered} IMF with \gammab$=1.3$). Age and metallicity are estimated by fitting \hbo\ and \mgfep\ simultaneously (see e.g. LB13). { Note} that the effect of \mgfe\ on \hbo\ is rather small (see V15), while \mgfep\ is independent of \mgfe~(\citealt{Thomas:04}; V15). A proxy for \mgfe\ is obtained from \mgb\ and \fem\ and then converted to \mgfe\ with a similar approach as described in LB13 and V15.
 \item[ ii. ] Same as at point { i}, but using EMILES base models with BaSTI isochrones (hereafter BASTI iT).
 \item[ iii. ] We fit simultaneously \hbo, \mgfep, \mgb, and \fem\ using $\alpha$--MILES SSP models.
 \item[ iv. ] Age and metallicity are obtained by averaging out results from different sets of  spectral indices, fitting simultaneously also the IMF slope (see Sec.~\ref{sec:IMFfitting} below).
\end{description}
{ Note} that we use different methods/models in order to account for possible systematic effects, and/or the effect of uncertainties on stellar population models.

\subsection{Constraining the stellar IMF}
\label{sec:IMFfitting}
We constrain the stellar IMF following the same approach as in LB19 (and references therein).   For each spectrum, we minimize the expression,
\begin{eqnarray}
\rm \chi^2(Age, {\rm [Z/H]}, \Gamma_b, [X/Fe]) = 
  \rm \sum_i \left[ \frac{E_{obs,i}- E_{mod,i} }{\sigma_{E_{obs,i}}} \right]^2 
\label{eq:method}
\end{eqnarray} 
where the  index $i$ runs over a selected  set of spectral features (see below); $\rm E_{mod,i}$ are line-strength predictions for Na--EMILES stellar population models; 
$\rm E_{obs,i}$ and $\sigma_{E_{obs,i}}$ are observed line-strengths and their uncertainties; \xfe\ are the elemental abundance ratios (e.g. \nafe ) included in the fitting procedure. For each fitting case (see below), uncertainties on best-fitting parameters, \{Age, ${\rm [Z/H]}$, \gammab, \xfe \}, are obtained from $N=1000$ bootstrap iterations, where the fitting is repeated after shifting observed line strengths according to their uncertainties.

In order to obtain robust results, we consider different fitting methods, by (i) changing  the set of fitted indices, (ii) changing the method to account for non-solar abundance ratios (see below); (iii) exploring either 1SSP or a combination of two SSP models, and/or imposing additional constraints to the (luminosity-weighted) age of the best-fitting model; (iv) using models based on different isochrones, i.e. BaSTI and Padova00 (see Sec.~\ref{sec:spmodels}). The options adopted for different methods are summarized in Tab.~\ref{tab:methods}. The rationale behind options (ii) and (iii) is the following:
\begin{description}
 \item[ Abundance ratios - ] In our ``reference'' approach (method A), we correct line-strengths to solar scale as a function of metallicity (see col.~6 of Tab.~\ref{tab:methods}). The corrections are based on observed trends of line-strengths, at fixed galaxy velocity dispersion, for SDSS ETGs' spectra.  As shown in LB16 and LB19, this approach was able to match well the observed line indices of massive ETGs at different galacto-centric distances. Since we found that the same approach does not allow us to match all observed indices for M31 (see below), we also consider different options (see col.~5 of Tab.~\ref{tab:methods}), where additional elemental abundance ratios, \xfe,  are included in the fitting procedure. To this effect, the sensitivity of a given index to \xfe\ is estimated from CvD12a stellar population model predictions, and an additional term is included in Eq.~\ref{eq:method} (see LB13 and LB15 for details). { Note} that this method  is not applied to \nafe, whose effect is already taken into account by Na--EMILES models. In general, when including a specific elemental abundance in the fitting, we also include indices with a prominent sensitivity to it~\footnote{For instance, in method E, we also include \cafe\ abundance ratio in the fitting procedure, since CaT lines are sensitive to both IMF and Ca abundance. The \cafe\ is constrained by adding \caii\ as well as \caf\ to the list of indices. Since \caf\ is  anticorrelated to \cfe, we include also \cfe\ as a further fitting parameter (see Tab.~\ref{tab:methods}), and constrain the \cfe\ by fitting also the \cfs\ band.}. { Note} that when applying empirical corrections and including \xfe's in the fitting procedure, one should interpret the \xfe's as ``residual'' abundance ratios (i.e. not accounted for by the empirical correction, as discussed in LB13).
 \item[ Constraints to age - ] As discussed in LB13, Balmer lines  have some sensitivity to IMF. The \hbo\ index decreases with IMF, so that for a bottom-heavier distribution one tends to infer younger ages, and vice versa. However, this IMF sensitivity turns out to be model dependent, as CvD12a models do not show a dependence of Balmer lines on IMF (see V15). Moreover, the \hbo\ decreases with \mgfe, and is sensitive to other abundance ratios (e.g. \cfe). In order to overcome these issues, for each spectrum of M31, we also derive a luminosity-weighted age estimate with spectral fitting, using the software {\sc pPXF}. For some fitting methods (see col.~7 of Tab.~\ref{tab:methods}), the {\sc pPXF} luminosity-weighted age is also added as extra constraint (i.e. an extra term) in Eq.~\ref{eq:method} (see eq.~1 of LB19).
 To explore the effect of different star-formation histories, we also use both 1SSP and 2SSP models, where the effect of a second young component is taken into account (see methods J and K in Tab.~\ref{tab:methods}).  
\end{description}

\begin{table*}
\centering
\small
 \caption{Summary of the different methods used to infer the IMF slope. Column 1 lists the labels used for different methods. Column 2 reports the number of SSPs in the fitting procedure, while column 3 refers to the models' isochrones,  with label ``iP'' (``iT'') for Padova00 (BaSTI) isochrones (see Sec.\ref{sec:spmodels}). Columns 4 and~5 give the list of indices and the elemental abundance ratios included in the fitting. Columns 6  and 7 flag the case where empirical corrections to line-strengths and age constraints from spectral fitting were applied (see the text).}
  \begin{tabular}{c|c|c|c|c|c|c}
   \hline
 Method &  \#SSPs & isochrones & Spectral indices & \xfe's & Empirical & Age \\
        &       & & & & corrections & constraints \\
   (1)  &     (2)      &  (3) & (4)   & (5) & (6) & (7)    \\
  \hline
    A &  1 & iP & \hbo, \mgfep, \mgf, \tioi & none & yes & yes \\
      &    &    & \tioii, \atio, \nad, \naii, \cat  &      &   &  \\
    B &  1 & iP & same as A w/o TiO's & none & yes & yes \\
    C &  1 & iP & same as A w/o CaT   & none & yes & yes \\
    D &  1 & iP &       same B        & none & yes &  no \\
    E &  1 & iP & same as A, \caii, \caf, \cfs & \cfe, \cafe & yes & yes \\
    F &  1 & iP & same as E w/o \tioii, \atio & \cfe, \cafe & yes & yes \\
    G &  1 & iT & same as E  & \cfe, \cafe & yes & yes \\
    H &  1 & iP & same as E w/o \hbo & \cfe, \cafe & yes & yes \\
    I &  1 & iP & \hbo, \mgfep, \tioi, \tioii, \atio, & \cafe, \tife, \ofe, \cfe, & yes & no \\
      &    &    & \mgf, \nad, \naii, \cai, & \nfe, \mgfe, \sife &     &     \\
      &    &    & \caii, \cat, \caf, \mgi, \mgii &  &     &     \\
      &    &    & \cfs, \cnii, \mgb, \feff &  &     &     \\
    J &  2 & iP &    same as B & none & yes & no \\
    K &  2 & iP &  same as I + \hgf & none & yes & yes \\
   \hline
  \end{tabular}
\label{tab:methods}
\end{table*}

\subsection{Best-fit spectral indices}
\label{sec:bestfit}

\begin{figure*}
 \begin{center}
\leavevmode
    \includegraphics[width=14cm]{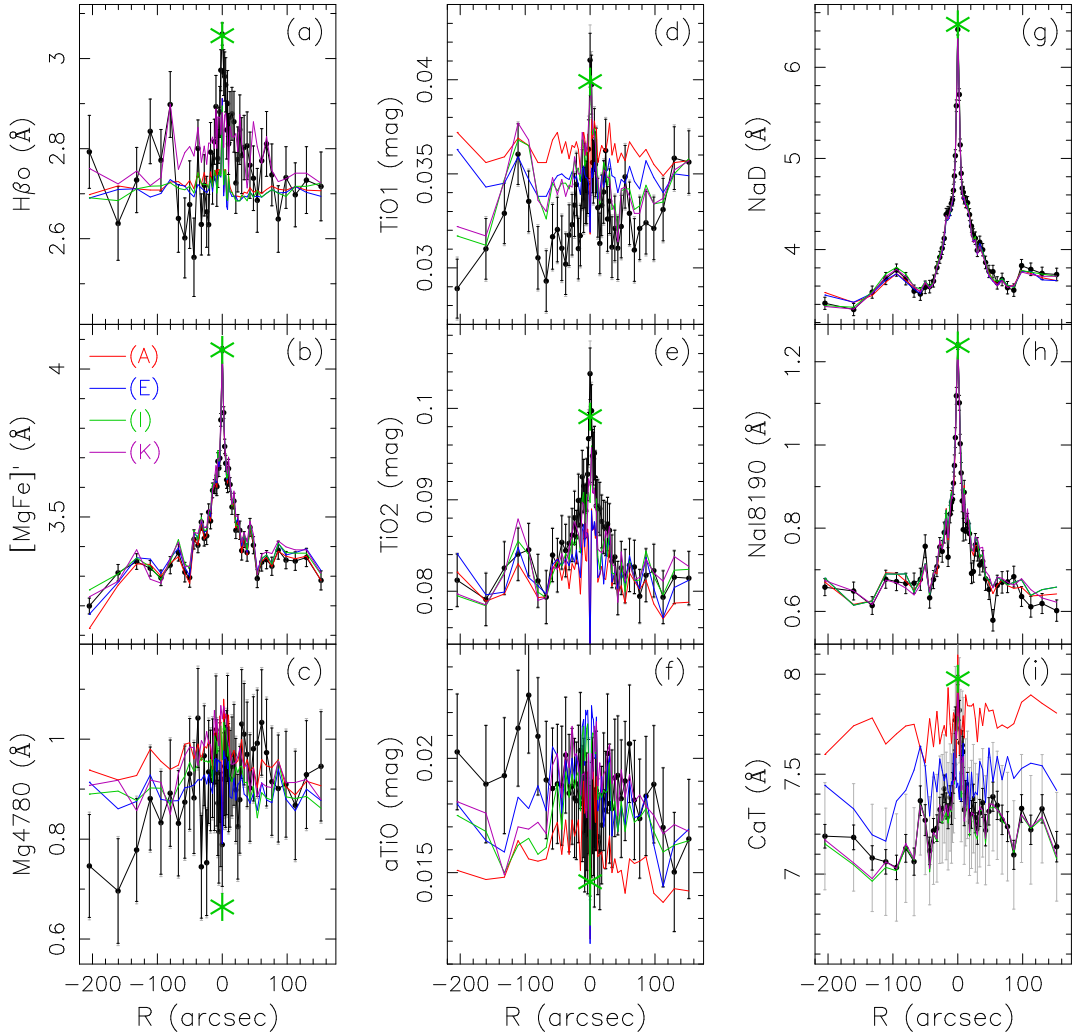}
 \end{center}
    \caption{Observed indices for the bulge of M31 as a function of galactocentric distance. Different panels correspond to different spectral indices. All indices (black dots with error bars) have been corrected to a sigma of 150\kms\ (see the text). Black error bars account for different sources of uncertainties on observed line-strengths, as detailed in Sec.~\ref{sec:analysis}. Grey error bars account for the uncertainty on the empirical correction (methods A and E; see the text). { Note} that grey error bars are not shown for \hbo, \mgfep , and Na indices, as no correction is applied to these indices. For all the other indices but \cat , grey error bars are almost indistinguishable from the black ones.  
    Solid lines with different colours show best-fitting results for a representative set of different fitting methods (i.e. methods A, E, I, and K, see labels in panel b,  and Sec.~\ref{subsec:IMFgradient}). The green asterisks show best-fitting results for the innermost radial bin when treating the dependence of \tioi\ and \tioii\ on metallicity as free fitting parameters (see the text). 
    }
    \label{fig:indices}
\end{figure*}

\begin{figure*}
 \begin{center}
\leavevmode
    \includegraphics[width=14cm]{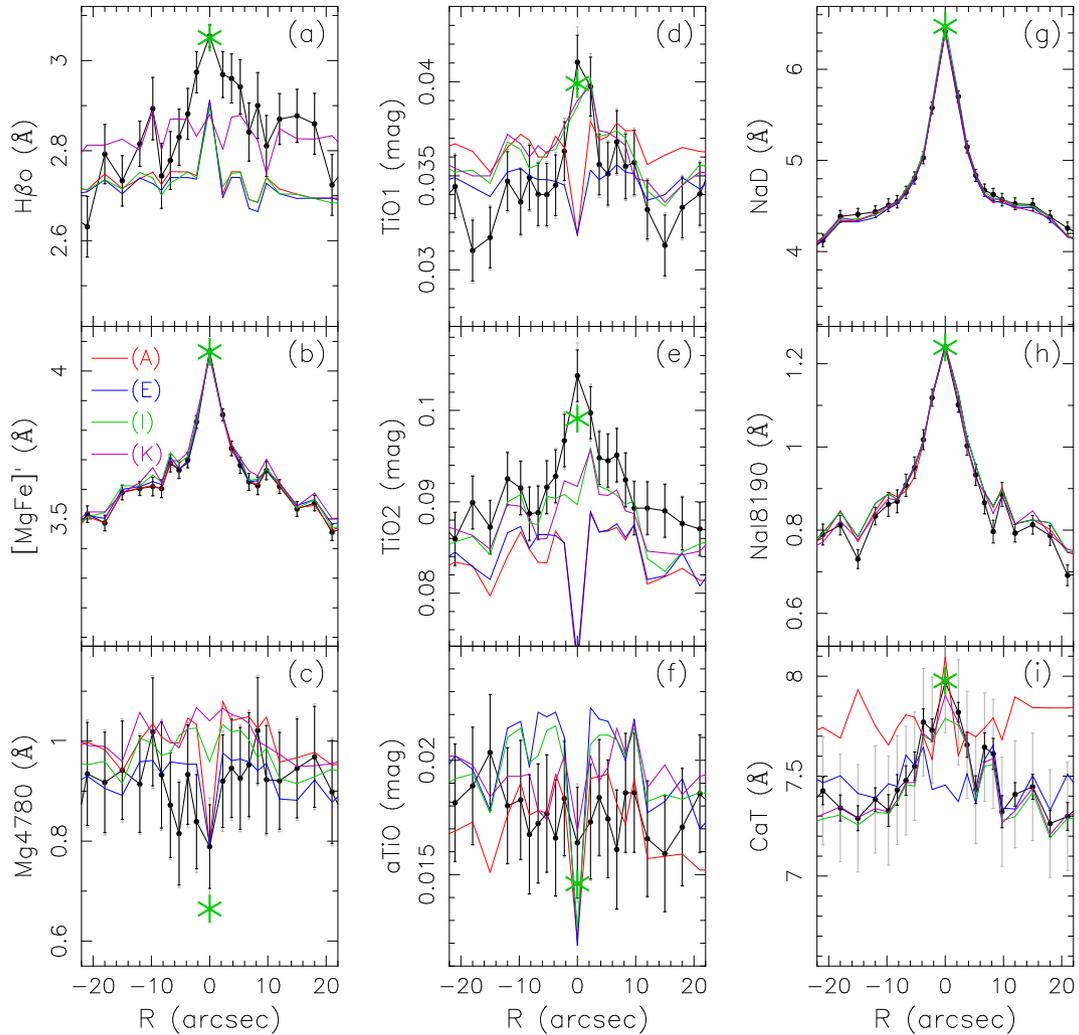}
 \end{center}
    \caption{Same as Fig.~\ref{fig:indices} but zooming into the inner region within $\sim 20$'' from the bulge center.
    }
    \label{fig:indices2}
\end{figure*}

Fig.~\ref{fig:indices} compares observed and best-fit line-strengths for the bulge of M31, as a function of galactocentric distance $\rm R$, in units of arcsec. Fig.~\ref{fig:indices2} shows the same comparison but zooming into an innermost region of about $\pm 20$'' from the center. To illustrate the quality of the fits, the Figures show a selected set of spectral indices, including all the IMF-sensitive features analyzed in this work, i.e. the TiO's (panels d and e), Na indices (panels g and h), \atio\ (panel f), \mgf\ (panel c), and the \cat\ (panel i),  as well as the age and metallicity indicators \hbo\ and \mgfep\, in panels a and b, respectively.
In order to compare different radial bins, all the line-strengths plotted in Figs.~\ref{fig:indices} and~\ref{fig:indices2} have been corrected~\footnote{{ Note} that this correction is only performed to display  observed and model indices for all radial bins in the same figure.  For each radial bin, we fit observed line-strengths with predictions of models smoothed at the sigma of the given bin.} to the same sigma~\footnote{The correction is estimated using results from method A. In a given radial bin, the SSP model corresponding to the best-fitting parameters (age, metallicity, IMF, and \xfe ) is smoothed at $\sigma=150$~\kms\ and at the actual sigma of the bin. The correction is performed by computing the difference of the model line-strengths measured for both sigma values, and adding this difference to the observed line-strength.
} of 150~\kms. Observed line-strengths, with error bars, are plotted in black, while  best-fitting indices are plotted with different colours, for a representative set of fitting methods from those listed in Tab.~\ref{tab:methods}.
\begin{description}
 \item[- ] Method A (red line) is our basic approach, where we fit all IMF sensitive features, together with \hbo\ and \mgfep, applying an empirical correction to line-strengths for the effect of abundance ratios (see Sec.~\ref{sec:IMFfitting}). Fig.~\ref{fig:indices} shows that most of the best-fitting indices (red line) are consistent with the observed ones within the error bars, although a significant offset is seen for \tioi, \atio, and \cat. For the TiO's and CaT, the best-fit does not match the rapid increase of observed indices in the bulge central region (Fig.~\ref{fig:indices2}).
 \item[-] Method E (blue line) shows the case where some residual abundance ratios, i.e. \cfe\ and \cafe, are included in the fitting.
 While including \cafe\ removes most of the offset for \cat\ (see above), as in the case of method A, the best-fit solution does not provide a good match of the radial trends in the innermost radial bins.
 \item[-] Method I (green line) fits all abundance ratios by combining Na--EMILES models with responses from CvD12a~\footnote{The effect of \nafe\ is already accounted for by Na--MILES. Hence, no response from CvD12a is used in this case.} (see Tab.~\ref{tab:methods}). We see that best-fitting  line-strengths match extremely well all the observed line-strengths, but for the innermost radial bin, where models (green curve) do not match the peak of \tioi\ and \tioii\ in the center (see Fig.~\ref{fig:indices2}). Method K (magenta curve) shows that a younger component (i.e. a 2SSP fit) does not improve the matching. 
\end{description}

In summary, Figs.~\ref{fig:indices} and~\ref{fig:indices2} show that we can fit reasonably well all observed indices of M31, at all radial position, but for the central bin, where we do not match the high values of \tioi\ and \tioii. As discussed in App.~\ref{app:tios}, the mismatch is likely due to the models' extrapolation in the very high metallicity regime, that predicts both { TiO indices} to be independent of metallicity up to \zh$\sim 0.4$ and above. Indeed, if we modify our best-fitting method I to allow for both \tioi\ and \tioii\ to increase with \zh\ (see App.~\ref{app:tios} for details), we are able to fit all observed line-strengths of M31 in the innermost bin, as shown by the the green asterisks in Figs.~\ref{fig:indices} and~\ref{fig:indices2}. { Some deviation remains only for \mgf\ in the central bin (see green asterisk in panel c of Fig.~\ref{fig:indices2}), though the discrepancy is not significant within the error bars ($< 1.5$~sigma level). { Note} also that \mgf\ shows some asymmetric behaviour with radius, with marginally lower values at negative $\rm R$,  though differences are not significant within the error bars~\footnote{
 These differences might be explained  by the fact that negative and positive values of $\rm R$ have been observed with different OSIRIS CCDs (see also Sec.~\ref{subsec:reduction} and Sec.~\ref{subsec:IMFgradient}).
}. }
For the present purposes, the key point is that, regardless of the fitting quality of individual methods, the fitting results, and in particular the IMF determination for M31, are very robust, as shown in the following sections.

% Since in the innermost bin, the best-fitting metallicity is above %the (safe) limit of EMILES SSP models (see Sec.~%\ref{sec:IMFfitting}), i.e. \zh$=0.26$, the best-fitting value are %affected by the models' extrapolation. Indeed, we assuming that both %\tioi\ and \tioii\ increase with metallicity, with a slope 
% Alternatively, some of the responses from CvD12 models might change %significantly with \zh, and thus account for the observed increase %of \tioi\ and \tioii\ in the center of M31.

\section{Results}
\label{sec:results}

\subsection{Age, metallicity, and \mgfe\ radial profiles}
\label{subsec:agemetafe}

\begin{figure*}
 \begin{center}
\leavevmode
    \includegraphics[width=14cm]{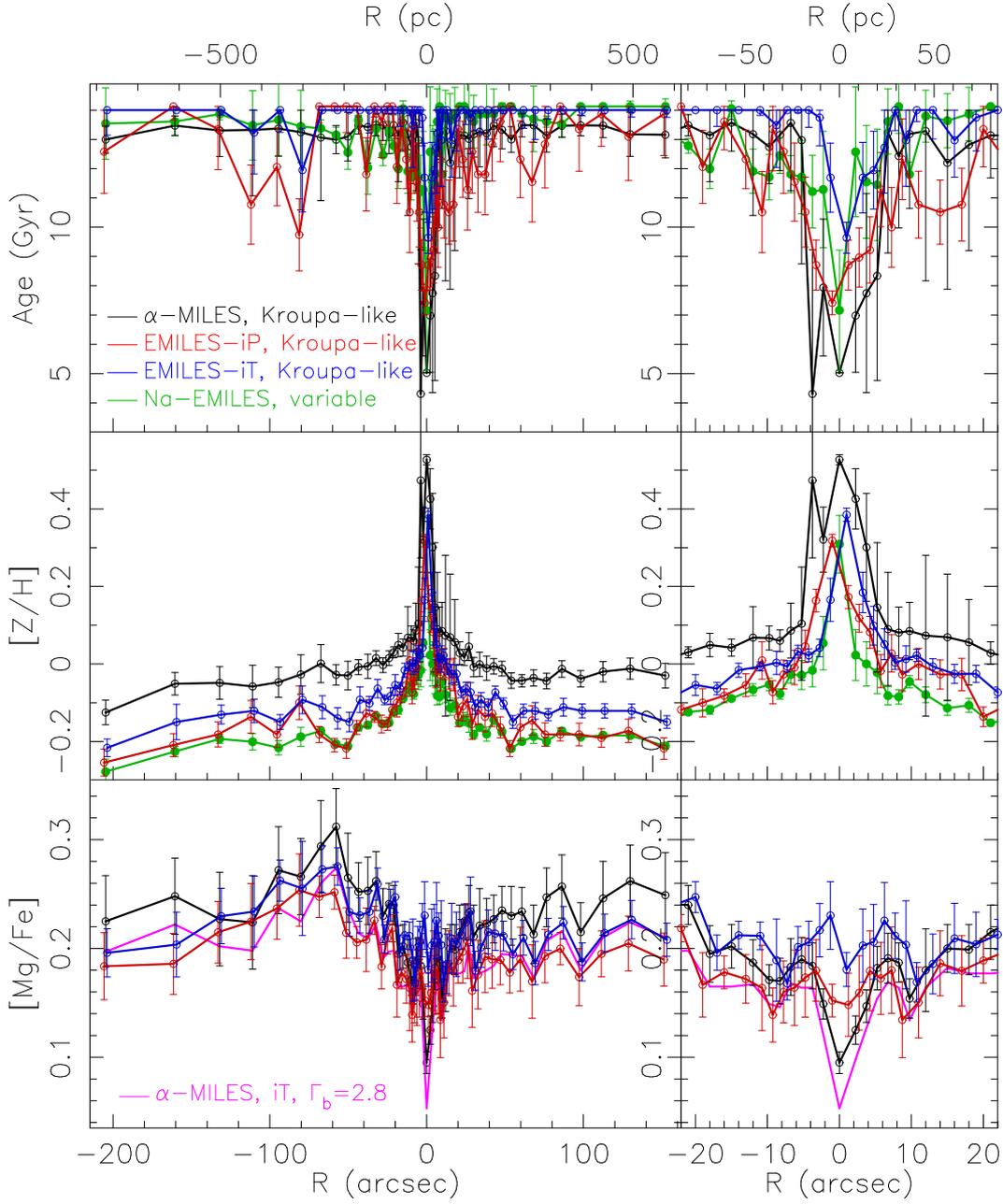}
 \end{center}
    \caption{Age (top), metallicity (middle), and \mgfe\ (bottom) as a function of galactocentric distance for the bulge of M31. The panels on the right zoom into the innermost region of the bulge, within $\sim 20$'' from the center.
    The bottom (top) axes show the radial distance $\rm R$, in units of arcsec ($\rm pc$).
    Black, red, blue, and green curves show results obtained with different stellar population models, as well as by averaging results from different methods to constrain the stellar IMF (green curve), as labeled in the top--left panel (see the text). In the bottom panels, magenta curves and symbols show \mgfe\ estimates obtained from $\alpha$--MILES models with a bottom-heavy (\gammab$=2.8$), rather than a Kroupa-like, IMF.
    Error bars denote 1-sigma uncertainties.
    }
    \label{fig:agemetalfa}
\end{figure*}

Fig.~\ref{fig:agemetalfa} plots the radial profiles of age, metallicity, and \mgfe\ for the bulge of M31, comparing the different methods we adopted to derive these quantities, as detailed in Sec.~\ref{sec:agez}. The right panels in the Figure zoom into the inner bulge region (within a radial distance of $\sim 20$'' from the center).

Our OSIRIS spectroscopy shows that the bulge of M31 is characterized by old stellar populations, with ages older than $\sim 11$~Gyr, at all radial positions, but for the innermost region, within $\sim 6$'' ($\sim 23$~pc), where the age  is as young as $\sim 5$--$7$~Gyr. For what concerns metallicity, some offset exists among results from different models. For radial distances $\gtrsim 20$'' from the center, metallicity is slightly subsolar (around solar), with \zh$\sim -0.2$ (\zh$\sim 0$) for EMILES ($\alpha$--MILES) models. In the innermost region (within $\sim 10$~arcsec), where  age is younger, metallicity is as high as $\sim 0.3$--$0.4$~dex (depending on the method).
The \mgfe\ is almost constant ($\sim 0.2$--$0.24$~dex),
at all radii probed in this work (out to $\sim 200$''), decreasing to $\sim ~0.1$--$0.15$~dex (depending on the method used to estimate \mgfe) only in the innermost radial bin.
The presence of a (mildly) positive \mgfe\ gradient is also confirmed by the analysis of the \mgb--Fe index-index diagram, as shown in App.~\ref{app:azagrids}, where we also discuss 
differences between results from EMILES and $\alpha$--MILES models, respectively.
The fact that the innermost radial bins are characterized by a young age and lower \mgfe\ points to a more extended star-formation history in the central region (within $\sim 20$~pc) of the bulge.
The magenta curves in the bottom panels of Fig.~\ref{fig:agemetalfa} show that the \mgfe\ is essentially independent of the assumed IMF, with a bottom-heavy distribution (\gammab$=2.8$) giving consistent results~\footnote{The fact that \mgfe\ is independent of IMF is relevant as we use the \mgfe\ estimates obtained from $\alpha$--MILES models with a Kroupa-like IMF, as input to our fitting procedure to constrain the stellar IMF (Sec.~\ref{sec:IMFfitting}).} to those for a Kroupa-like IMF, in agreement with what already found by LB17. 

\subsection{The IMF radial gradient}
\label{subsec:IMFgradient}

\begin{figure*}
 \begin{center}
\leavevmode
    \includegraphics[width=16cm]{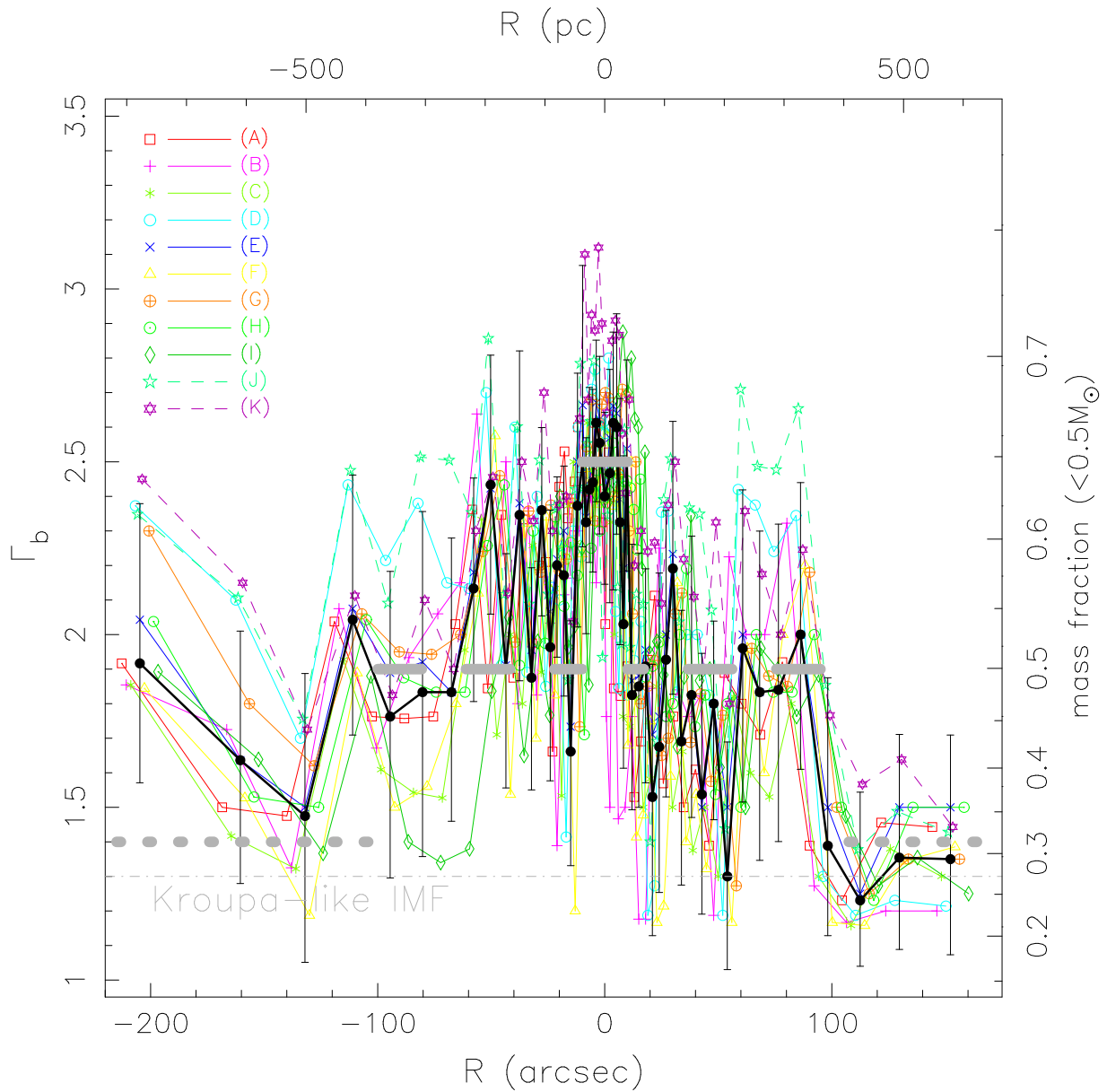}
 \end{center}
    \caption{The IMF slope, \gammab, for the bulge of M31 is plotted 
    as a function of the radial distance $\rm R$, in units of arcsec ($\rm pc$), along the bottom (top) horizontal axis.
    Negative and positive values of $R$ refer to opposite directions along the OSIRIS slit (see Sec.~\ref{subsec:binning}). Different colours are the IMF profiles obtained with different fitting methods to spectral indices (see labels in the top--left of the Figure and Tab.~\ref{tab:methods}).
    The median-combined profile of \gammab\ is shown with a black thick line and circles with error bars, corresponding to 1--sigma standard errors. The horizontal grey dot-dashed line marks the value of \gammab$=1.3$, for which the { low-mass tapered} IMF approximates the Kroupa IMF. Dotted, dashed, and solid thick segments, plotted in grey, define different radial regions, where the IMF is consistent with a Kroupa-like distribution ($100 \le |R| \le 200$''), marginally above it ($10 \le |R| \le 100$''), mildly bottom-heavy ($|R|<10$''), respectively (see the text). The y-axis on the right side of the plot shows the mass fraction of low-mass (<$0.5 \, M_\odot$) stars in the IMF, as a function of \gammab, as defined in LB13.
    }
    \label{fig:gammab}
\end{figure*}

Fig.~\ref{fig:gammab} shows the IMF radial profile for the bulge of  M31. Results from different methods to infer the IMF slope, \gammab , are shown with different colours (as labeled in the top--left of the Figure; see Sec.~\ref{sec:IMFfitting} and Tab.~\ref{tab:methods}). The final IMF profile is obtained by averaging results from different methods, and is shown as a thick black curve with error bars. We find a mild, but significant, IMF radial gradient for the bulge of M31, with \gammab\ being consistent with a Kroupa-like distribution (\gammab$\sim 1.3$) in the outermost regions probed by OSIRIS ($\rm  R \gtrsim 100$''), and a ``mildly'' bottom-heavy distribution (\gammab$\sim2.5$) in the central bins. For comparison, massive ETGs can have a value of \gammab\ as high as $3$, or even higher, in their central regions (see Sec.~\ref{sec:Discussion}). { Note} that although some offset exists among different methods to infer the IMF, likely because of different systematics/model uncertainties when using different sets of spectral indices, the IMF radial trend is qualitatively the same for all methods.
This points to the importance of using a wide set of absorption features, from different chemical species, when constraining the IMF (see LB13 and~LB17).
We { note} that also  modifying method I to allow for a possible dependence of both TiO indices on \zh\ (see App.~\ref{app:tios}), does not change significantly our conclusions. In this case, we get an IMF slope of \gammab$\sim 2.3$ at $\rm R=0$, compared to  \gammab$\sim 2.4$ for method I (see Fig.~\ref{fig:gammab}). 

The IMF profile of M31 suggests the presence of three different radial regions, marked  with  grey thick segments in Fig.~\ref{fig:gammab}, i.e. (i) the radial range from $\sim 100$ to $\sim 200$'', with an average value of \gammab$\sim 1.4$ (see the grey dotted thick segments), where the IMF slope is fully consistent with a Kroupa-like distribution (i.e. \gammab$\sim 1.3$, plotted as a grey horizontal  dot-dashed line); (ii) the range from $\sim 10$ to $\sim 100$'', where the IMF is marginally above a Kroupa-like distribution, with \gammab$\sim 1.8$ (see grey dashed thick segments); and (iii) the innermost region, with $\rm |R| < 10$'', where the IMF is mildly bottom-heavy, with \gammab$\sim 2.5$ (see the grey solid thick segment). { Note} that the IMF profile appears to be slightly asymmetric, with values of \gammab\ higher for negative (relative to positive) values of $\rm R$. The effect is more pronounced at larger galactocentric distances. For $\rm R \gtrsim 100$'', the median value of \gammab\ is $\sim 1.3$, while for $\rm R \lesssim -100$'', we obtain a median value of \gammab$\sim 1.75$. We  { note} that the difference is small, given the quoted error bars, and mimics the small asymmetry seen in some of the observed line-strength profiles. In particular, as shown in Fig.~\ref{fig:indices}, while the radial profile of \naii\ is symmetric, the \nad\ tends to be lower for $\rm R<-100$'', hence favouring lower \nafe, and thus higher \gammab, than for $\rm R>100$''. This behaviour might be due to (i) intrinsic small differences in the stellar IMF among the two sides of the slit; (ii) a result of some residual contamination of NaD from interstellar absorption (see App.~\ref{app:ism}); (iii) the fact that negative and positive values of $\rm R$ are observed with different OSIRIS CCDs, and thus may be affected by different (small) systematic effects (e.g. flux calibration). { Note} that some (small) asymmetries are also seen in the radial profiles of \tioi, \atio, and \mgf. While it is hard to establish if such asymmetries are real, or result from data reduction, this issue does not affect at all the main conclusions of our analysis. Also, we emphasize that it is far more difficult to constrain the IMF slope when it is close to $\sim 1.3$ compared to the range with, e.g., \gammab$\gtrsim 2.3$, as indices have a much weaker dependence on IMF slope for low \gammab\ values (see e.g. LB13).

\begin{figure}
 \begin{center}
\leavevmode
    \includegraphics[width=8.5cm]{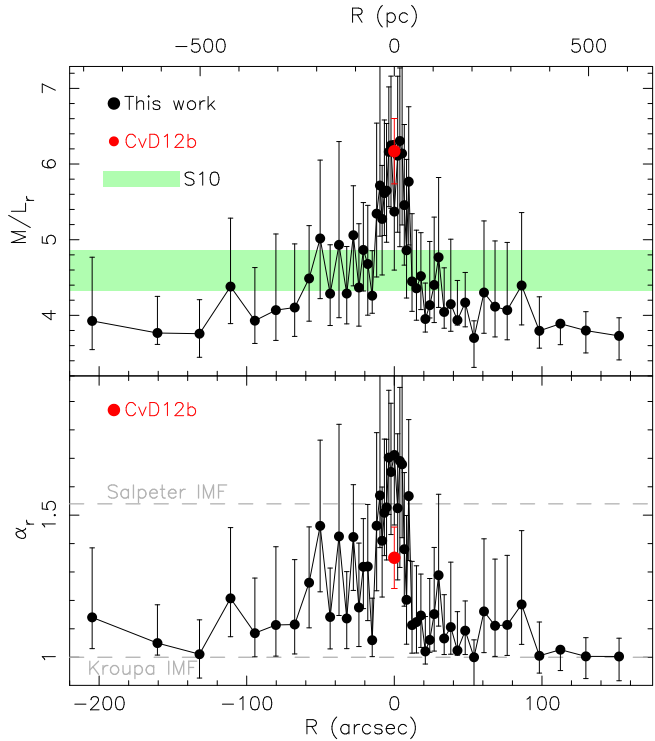}
 \end{center}
    \caption{Stellar mass-to-light ratio, \mlr\ (top), and mass-excess factor, $\rm \alpha_r$ (bottom), for the bulge of M31, as a function of galactocentric distance (i.e. the same as in Fig.~\ref{fig:gammab} for IMF slope). The $\rm \alpha_r$ is obtained by normalizing the \mlr\ by the expected value for a Kroupa-like IMF (see the text). The black circles with error bars (1-sigma confidence levels) are obtained by averaging results from different fitting methods (Sec.~\ref{sec:IMFfitting}). 
    The red dots correspond to estimates obtained by CvD12b for the nuclear region (within a radius of 4''), while the light-green region in the top panel marks the range of values for \ml\ found by~S10, under the assumption of a fixed Kroupa IMF (see Sec.~\ref{subsec:previousworks}). In the bottom panel, the dashed grey horizontal lines mark the values of $\rm \alpha_r$ for a Kroupa ($=1$) and Salpeter ($=1.54$) IMF, as labeled.
    }
    \label{fig:mlr}
\end{figure}

Fig.~\ref{fig:mlr} (top) plots the stellar mass-to-light ratio in r band, \mlr, as a function of radial distance to the center of the M31 bulge (see black dots with error bars). For each radial bin, the \mlr\ is obtained by averaging results from all different fitting methods (see above). The \mlr\ is approximately flat outside $60$'', with \mlr$\sim 3.9$, increasing up to $\sim 6$ in the innermost few arcsec region. Such an increase results from the radial variation of age, metallicity, and IMF. To single out the IMF effect, the bottom panel of Fig.~\ref{fig:mlr} plots the ``mass-excess'' factor, $\rm \alpha_r$, i.e. the actual \mlr\ normalized to that predicted for a Kroupa-like IMF, keeping the other stellar population  parameters (i.e. age and metallicity) unchanged. A value of $\rm \alpha_r \sim 1$ ($\sim 1.54$) corresponds to a Kroupa (Salpeter) IMF, as shown by the horizontal grey dashed lines in the bottom panel.
Overall, the radial behaviour of $\rm \alpha_r$ reflects that of \gammab, with a mildly bottom-heavy IMF (Salpeter, or ``slightly'' above Salpeter), in the innermost few arcsec, and a distribution consistent with a Kroupa-like IMF in the outer radial bins. { Note} that from \gammab$= 1.3$ to $1.8$, the \mlr\ increases by only $\sim 12$\% (assuming an SSP model with age of 12~Gyr and solar metallicity). Hence, it is difficult to define different radial ranges for \mlr\ and $\rm \alpha_r$ as for \gammab\ (see the grey segments in Fig.~\ref{fig:gammab}). 

\subsection{[Na/Fe] radial gradients}
\label{subsec:NaFegradients}

\begin{figure}
 \begin{center}
\leavevmode
    \includegraphics[width=8cm]{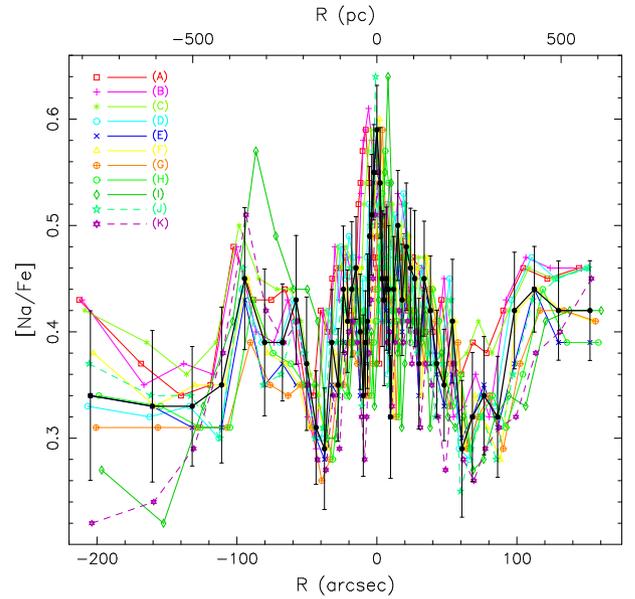}
 \end{center}
    \caption{The same as Fig.~\ref{fig:gammab} but plotting \nafe\ abundance ratio, rather than IMF slope. 
    }
    \label{fig:NAFE_M31}
\end{figure}

Since we rely on Na--enhanced EMILES stellar population models, our fitting procedure provides also an estimate of \nafe\ at different radial positions. The \nafe\ radial profile for the bulge of M31 is shown in Fig.~\ref{fig:NAFE_M31}, with different methods plotted with different colours, and the combined \nafe\ shown in black, as for \gammab\ in Fig.~\ref{fig:gammab}. Overall, the \nafe\ profile of M31 tends to be rather flat at all positions, but for the innermost region ($\lesssim 5$''). For $R>5$'', the median value of \nafe\ amounts to $0.41 \pm 0.06$~dex, where the error bar reflects the scatter among different radial bins. The maximum value of \nafe\ is reached in the central bin, with \nafe$=0.59 \pm 0.05$~dex.
For comparison, massive ETGs can have \nafe\ in the range $0.6$--$0.9$~dex, in their central regions (see, e.g., LB19; and Sec.~\ref{sec:Discussion}). We point out that, as discussed in LB17, the effect of \nafe\ and IMF are coupled in Na--MILES models, so that the effect of \nafe\ on Na indices is stronger for a bottom-heavy, relative to a Kroupa-like, distribution. This allows us to match the strong Na line-strengths of M31, in the innermost regions, without requiring a very  bottom-heavy IMF, nor an extremely high \nafe .

\section{Discussion}
\label{sec:Discussion}

\subsection{Age, metallicity, and \afe}
\label{subsec:previousworks}

Our  age and metallicity radial profiles (Fig.~\ref{fig:agemetalfa}) are qualitatively consistent with those of S10 (see also~\citealt{Saglia:2018}), who found that, overall, the bulge of M31 is characterized by  old ages ($>10$~Gyr) and solar metallicity, with younger ages ($\sim 8$--10~Gyr) and supersolar metallicities (up to \zh$\sim$0.4~dex) at galactocentric distances smaller than a few arcsec (see fig.~12 of S10).  We 
also find a small, positive, radial gradient of \mgfe\ (see App.~\ref{app:azagrids}), not detected by S10, likely because of the  uncertainties on \afe\ (see, e.g., their fig.~12). Since the \mgfe\ can be interpreted as a proxy of the star-formation timescale in a stellar system~(\citealt{Thomas:05, deLaRosa:2011}; but see~\citealt{Fontanot:2017, Fontanot:2018}), the presence of young ages and lower \mgfe\ in the innermost radial bins of M31 consistently point to a more extended/recent star-formation in the bulge center. 
For the first time,  we also find a (mild) radial variation of the stellar IMF in the bulge of M31, with a  Salpeter-like ($\rm \alpha_r \sim 1.7 \pm 0.2$) IMF normalization (Fig.~\ref{fig:mlr}) within the innermost few arcsec, decreasing to a Kroupa-like normalization at larger radii (see Sec.~\ref{subsec:IMFgradient}).  

Our results for the central radial bins of M31 can be compared to those of~CvD12b, who constrained the stellar IMF within a region of $4$'' from the center of M31, 
using a different set of models, and a different approach (i.e. full spectral fitting, rather than index fitting). CvD12b found an IMF normalization between Kroupa and Salpeter for the nuclear region of M31. 
Fig.~\ref{fig:mlr} plots, as red dots, the \mlr\ and $\rm \alpha_r$  derived by CvD12b.
Our $\rm \alpha_r$ tends to be higher, and only marginally consistent (at about 2~sigma level), with that of CvD12b. Nonetheless, the \mlr\ is consistent with that of CvD12b, most likely because of the IMF parametrization adopted in the present work (i.e. a {low-mass tapered} distribution, rather than a multi-component power-law, as in CvD12b), as well as the young age (implying lower \mlr) that we estimate in the innermost radial bins.  CvD12b found also an extremely high \nafe\ abundance ratio ($\sim 1$~dex), in disagreement with our results, of \nafe\  $\sim 0.4$~dex, rising up to $\sim 0.6$~dex (at most) in the inner few arcsec (see Sec.~\ref{subsec:NaFegradients}). 
{ Note} that the data used by CvD12b for M31 did not include the \nad\ feature, which is crucial, indeed, to estimate \nafe\ and thus disentangle the effect of \nafe\ and IMF on Na features (see LB17).

\subsection{Mass-to-light ratios}
\label{subsec:ML}

S10 analysed spectral indices for the bulge of M31, using stellar population models by~\citet{Maraston05}. Under the assumption of a constant Kroupa IMF, they predicted R-band M/L ratios  in the range of 4 to 4.5. This range, corresponding to $\rm 4.3 \lesssim M/L_r \lesssim 4.9$~\footnote{Based on EMILES SSP models, we adopt a conversion factor of $\rm  (M/L_R) / (M/L_r) \sim 0.925$. }, is shown as a green shaded region in Fig.~\ref{fig:mlr}  (top panel). 
To perform a more direct comparison, we also computed from our M/L profile, under the assumption of circular symmetry, the luminosity-weighted value of $\rm M/L_r$ for the entire bulge, out to the maximum galactocentric distance ($\sim 200$'') covered by our analysis.
We find $\rm  \langle M/L_r \rangle =4.1^{+0.6}_{-0.4}$. 
The predictions of S10 are fully consistent with our $\rm  \langle M/L_r \rangle $ estimate. However, since we find a radially varying  IMF, our M/L's change significantly with radius, from $\sim 3.8$ (at R$\sim 200$'') up to $\sim 6$ (in the innermost radial bins). { Note} that \citet{Saglia:2018} also found, under the assumption of a Kroupa IMF, V-band M/L ratios in the range of 4.5 to 5 (see, e.g., their figs.~21--22), corresponding to $3.9 \lesssim M/L_r \lesssim 4.4$, still consistent with our $\rm  \langle M/L_r \rangle $, and consistent with our M/L estimates for galactocentric distances $\gtrsim 10''$.  S10 concluded that there is a good agreement between their estimated M/L and those from the dynamical model of~\citet[hereafter W03]{Widrow:2003}, 
with $\rm M/L_R \sim 4.5$ (i.e. $\rm M/L_r \sim 4.9$). 
{ Note} again that this value is consistent with our average M/L. However, since the dynamical models of W03 assume a radially constant M/L, the comparison should be taken with some caution. The same applies when comparing our findings to those of~\citet{Diaz:2018}. The authors constructed a detailed model of the M31 bulge,  including the contribution from different dynamical structures, such as a ``classical'' bulge, a boxy/peanut bulge, and a thin bar, concluding that the M/L of the two (classical and boxy/peanut) bulge components  is consistent with that expected for a Chabrier IMF ($\rm M/L_r \sim 4$), under the assumption of a radially constant M/L. 
{ Note} that, indeed, our $\rm  \langle M/L_r \rangle $ is very similar to that one would infer for a  Kroupa-like IMF (i.e. $M/L_r \sim 4$), as the innermost bins, { where we infer a bottom-heavier IMF}, give only a minor contribution to the integrated light (mass) of the bulge. 

\citet{Dutton:2013} used gravitational lensing to constrain the stellar IMF normalization (i.e. the stellar M/L) in the bulges of massive spiral galaxies, finding evidence for a Salpeter-like normalization. We point out that this result is not in disagreement with our findings for the bulge of M31, as the bulges analyzed by \citet{Dutton:2013} are in the velocity dispersion range of $200$--$250$~\kms , while the sigma of the M31 bulge is $150$--$160$~\kms. Considering the relation between IMF slope and velocity dispersion of ETGs, for a velocity dispersion of 160~\kms, the expected $M/L_r$ is  $3.5 \pm 1$ (for a { low-mass tapered} IMF; see fig.~12 of LB13), fully consistent with our luminosity-weighted estimate for the bulge of M31.  

The presence of a radially varying IMF, and thus the increase of $\rm M/L_r$ in the center of the bulge,  might also affect dynamical models of the M31 nucleus (at $R < 1$''), consisting of an eccentric stellar disk orbiting a central black hole (BH). Assuming a negligible contribution ($<1 \%$) of the bulge component to the nucleus, \citet{PT:2003} performed a detailed dynamical model, obtaining a BH mass estimate of $\sim 1.0 \times 10^8$~$\rm M_\odot$, by a factor of $\sim 2$ larger (but still within the uncertainties) compared to the estimate of $6 \times 10^7 \, \rm M_\odot$ from the correlation of BH mass and bulge dispersion of nearby galaxies~\citep{Tremaine:2002}. Indeed, an increase of \mlr\ by a factor of $\sim 2$ in the nucleus, as predicted by our results (Fig.~\ref{fig:mlr}), would still imply a negligible contribution of the bulge mass to the nuclear region, and thus not decrease significantly the BH mass estimate.  

Indeed, dynamical models allowing for radial M/L variations would help to further scrutinize our results, and test if radial IMF variations  are fully consistent with the bulge kinematics, and how/if they affect  the inferred properties (such as the BH mass) of the M31 nucleus. We postpone this analysis to a forthcoming contribution.

\subsection{Bottom-heavy or Na-enhanced?}
\label{subsec:IMF_NA}

The radial behaviour of IMF-sensitive features for the bulge of M31 has been also analyzed by~\citet[hereafter Z15]{Z:15}, who found a negative radial gradient for \naii\ and \cat, decreasing from a radial distance of a few arcsec out to a few hundred arcsec, in agreement with our results (see panels h and i in Fig.~\ref{fig:indices}). Z15 compared the radial behaviour of \naii\ and \nad\  for M31, with SSP model predictions from~\citet[hereafter V12]{Vazdekis:12} (i.e. an earlier version of EMILES models) and CvD12a model predictions. 
They concluded that CvD12a models predict a strong dependence of 
\naii\ on IMF, compared to \nad, while for V12 models, also \nad\ 
depends significantly on IMF (see also LB13). For this reason, CvD12a models would favour a MW-like IMF at all radial positions in M31, with an extreme \nafe\ abundance ratio in the center (up to 1~dex); while V12 models would suggest a bottom-heavy IMF  in this region, and a Salpeter-like IMF for the rest of the bulge. We { note} that the comparison performed by Z15 suffered from the lack of  predictions  with varying \nafe\ for V12 models, and relied on \naii\ and \nad\ measurements drawn from different sources (with \nad\ 
from~\citealt{Davidge:1991}, ~\citealt{Davidge:1997}). 
%We also notice that Z15 did not perform any fitting of M31 line-%strengths with model predictions. 
Indeed, using Na-enhanced stellar population models, and an homogeneous spectroscopic data-set, our analysis shows that for most of the bulge radial extent, all optical and NIR spectral features of M31 are well matched with a Kroupa-like IMF, and an \nafe\ abundance ratio of $\sim 0.4$~dex.

%S10 compared colours predicted from stellar population models %matching the spectral indices with observed, extinction-corrected %SDSS colours of M31 ($\rm u-g$, 
%$\rm g-r$, and $\rm r-i$), finding a good agreement.
%Indeed, since an enhanced fraction of low-mass stars (i.e. a %bottom-heavy IMF), implies a redder colour with respect to that for %a Kroupa-like distribution, a question arises if our results %predict colours in the bulge center that are consistent with %observations.

\subsection{M31 bulge vs. massive ETGs}
\label{subsec:IMF_M31_ETGs}

Based on the \tioii\ spectral index, ~\citet[hereafter MN15b]{NMN:15b} found a tight correlation between \zh\ and IMF slope for ETGs in the CALIFA spectroscopic survey~\citep{CALIFA:2012}, pointing to stellar metallicity as a possible (local) driver of IMF variations in massive ETGs, in agreement with what also found by~\citet{vanDokkum:2017}.
However, in LB19, we showed that very massive ETGs, mostly brightest 
cluster galaxies (BCGs), do not follow the IMF-metallicity relation. 
\citet{NMN:19} also showed that in a disk-like structure, IMF variations do not follow metallicity changes. Hence, metallicity does not seem to be the only culprit of IMF variations in galaxies. 
Our results for M31 further support this conclusion. Indeed, in the bulge innermost radial bins, where stellar metallicity  is very high (up to \zh$\sim 0.4$), we find only a mildly bottom-heavy IMF (\gammab$\sim 2.5$; see Fig.~\ref{fig:gammab}), while based on the IMF-metallicity relation (see, e.g., fig.~2 of MN15b), one would expect \gammab$\sim 3$ or above. { Note} that based on \tioi\ and \tioii, one would actually infer a very bottom-heavy IMF also for the innermost bins of M31, as these indices increase very steeply in the innermost few arcsec (see Fig.~\ref{fig:indices2}). However, as discussed in App.~\ref{app:tios}, the increase of \tioii\ (and \tioi ) with \zh\ in the center of M31, is most likely due to a genuine dependence of the TiO's on metallicity in the high-metallicity regime, rather than an IMF variation.
This highlights the importance of using a wide set of spectral features, throughout a large wavelength baseline, when constraining the IMF.

\begin{figure}
 \begin{center}
\leavevmode
    \includegraphics[width=9cm]{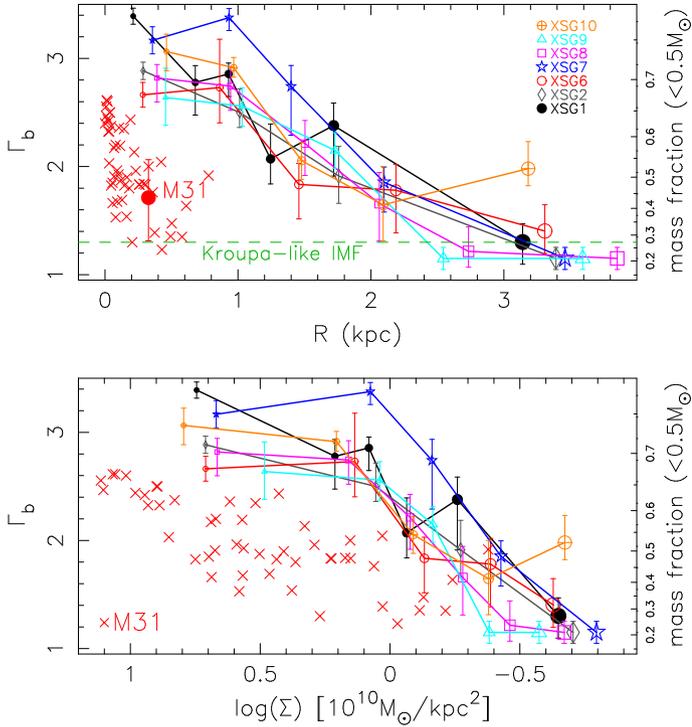}
 \end{center}
    \caption{The IMF slope, \gammab, of M31 is compared to the IMF radial profiles of massive ETGs, as a function of radial distance to galactocentric distance (top), and logarithmic stellar mass density (bottom).  In both panels, red crosses plot 
    the IMF slope values for different radial bins of M31, while
    other symbols, with different colours, plot the IMF radial profiles for the massive ``XSG'' ETGs from~LB19 (see labels on the upper--right). In the top panel, the red dot with error bars is the expected M31 IMF slope obtained by mimicking a circular aperture with average radius of 0.3~kpc, including the contribution from all radial bins (see the text for details). { Note} that the circularized value for M31 can be compared to the the innermost radial points of massive ETGs. The y-axis on the right of each panel shows the fraction of low-mass (<$0.5 \, M_\odot$) stars corresponding to a given value of \gammab\ (see LB13).
    In the top panel, the slope value for a Kroupa-like IMF is marked by the green horizontal dashed line.
    }
    \label{fig:M31_ETGs}
\end{figure}

In order to gain further insights into the origin of IMF variations, in Fig.~\ref{fig:M31_ETGs} we compare the estimates of \gammab\ for the bulge of M31 with those for very massive ETGs from LB19, as a function of galactocentric distance (top panel), and logarithmic stellar mass density ($\rm log(\Sigma)$; see bottom panel). The data for M31 (red crosses) cover a region out to $\sim 0.8$~kpc, and thus can be compared only with the innermost data-points for galaxies in LB19. To perform a direct comparison, we have computed, under the assumption of circular symmetry, the luminosity-weighted value of \gammab\ for the entire bulge, out to the maximum galactocentric distance ($r \sim 200$'') covered by our analysis. The red dot in the top panel of the Figure, with error bars, shows the luminosity-weighted IMF slope, i.e. \gammab$=1.7 \pm 0.4$.  { Note} that the integrated IMF slope is only marginally above the Kroupa-like value of \gammab$=1.3$, as the innermost bins, { where we infer a bottom-heavier IMF}, give only a minor contribution to the integrated light (mass) of the bulge. 
Fig.~\ref{fig:M31_ETGs} shows that the integrated IMF slope for the bulge is significantly lower compared to that for massive ETGs, within the same galactocentric distance. Hence, observing a spiral galaxy like M31 with the same spatial resolution as for ETGs, one would infer an IMF very similar to a Kroupa-like distribution. Also, although metallicity is very high in the center of the bulge, the IMF slope is $\sim 2.5$, while all massive ETGs tend to have higher \gammab ($>2.6$). 
{ Note} also that different radial bins for the M31 bulge cover a wide range in stellar mass density (see bottom panel of the Figure), allowing a direct comparison with different radial bins for massive ETGs. At fixed local density, the M31 spectra reveal a significantly lower \gammab\ compared to ETGs.
 The reason why the IMF is not very bottom-heavy in the center of M31, despite the high metallicity, might be its ``low'' velocity dispersion ($\sim 160$~\kms), compared to that of massive galaxies ($\sim 300$~\kms), or likely related to the formation mechanism itself of the bulge. From the relation between IMF slope and galaxy velocity dispersion of LB13 (see their fig.~12), ETGs with $\sim 160$~\kms\ are expected to have an ``integrated'' \gammab\ consistent with a Kroupa-like distribution, as we observe, indeed, for the bulge of M31.
 { Note} that, as shown in~\citet[see their fig.~3]{Tamm:2012},  the  nucleus gives a prominent contribution to the M31 surface brightness profile only at radial distances <$1--2$'', while we find IMF radial variations over a larger radial range ($<10$''; see, e.g., Fig.~\ref{fig:gammab}). Hence, the IMF variations are not driven by the nucleus of M31, but do actually reflect the properties of the bulge itself.
Indeed, as shown in~\citet{Bagna:2017}, the bulge of M31 consists of a ``classical'' and a ``box/peanut'' (BP) component, the latter contributing by $\sim$ 2/3 of the bulge total stellar mass.  The BP component is believed to form later from the disk material (possibly with a ``normal'' IMF), through the buckling instability of the bar. Therefore, one may speculate that any genuine IMF variation in the classical bulge might have been actually diluted by the stellar material formed in the BP component. Detailed dynamical models, including the effect of a varying IMF, might help to address this point. The comparison shown in Fig.~\ref{fig:M31_ETGs} 
reinforces the idea that there is no single parameter, such as stellar metallicity or local density, able to explain IMF variations in stellar systems~(see LB19).

\section{Summary and Conclusions}
\label{sec:Conclusions}
We have acquired new, long-slit spectroscopy with OSIRIS at GTC, along the major axis of the bulge of M31, obtaining an homogeneous, continuous spectral coverage in the wavelength range from $\sim 0.35$ to $\sim 1 \mu$m, with a constant spectral resolution of $\sim 38$~\kms (FWHM). Our dataset is the most homogeneous  obtained so far (in terms of spectral coverage and resolution) along the bulge major axis.
We fit a wide set of spectral indices, including Ca, Na, Mg, and TiO spectral features, with state-of-the-art stellar population models (Na--MILES), that take into account the coupled effect of varying IMF and Na abundance ratios.

Our analysis shows that the bulge of M31 has an IMF consistent with a Kroupa-like, or ``slightly'' above Kroupa, IMF (with IMF slope \gammab$<2$) at all radial positions but in the innermost 10'' (corresponding to ~$40$~pc), where a mildly bottom-heavy IMF (with a normalization close to the Salpeter distribution) is required. We are able to match both the \nad\ and \naii\ spectral indices simultaneously, without requiring extremely high \nafe, in contrast with previous studies. The best-fitting \nafe\ is $\sim 0.4$~dex, rising up to $\sim 0.6$~dex, in the inner few arcsec. 
The Na--enhanced models allow us to fit extremely well all IMF-sensitive spectral indices of M31, as well as a large set of Lick-like indices, once the effect of other individual abundance ratios is taken into account (using CvD12a SSP models). Compared to very massive ETGs (with a velocity dispersion $\gtrsim 300$~\kms), the bulge of M31 has a less bottom-heavy IMF at fixed stellar mass density, and a less bottom-heavy IMF in the center, despite having similarly high stellar metallicity. Significant IMF variations in the bulge are concentrated within a much smaller region ($< 100$~pc) compared to most massive ETGs, where the IMF changes occur at galactocentric distances larger than $\sim 1$~kpc. 
The integrated, luminosity-weighted, IMF slope is consistent with that expected for a Milky-Way-like distribution, and consistent with lower-mass ETGs, having the same velocity dispersion of the bulge ($\sigma \sim 150$--$160$~\kms ). The IMF radial gradient of the M31 bulge might originate from the presence of different dynamical structures (such as a classical and a box/peanut bulge), and IMF radial gradients might actually help to constrain the formation and evolution scenarios of different sub-systems. In the future, this point should be addressed with detailed dynamical models, including the effect of a radially varying  IMF.
Our results reinforce the idea that there is no single driver of IMF variations in galaxies, and we cannot entirely ascribe IMF variations to a single parameter, such as stellar metallicity, or observed mass density. The IMF gradients we observe at $z \sim 0$ likely result from a complex multi-component physical process, related to the details of the physical properties of the gas at the time of formation,  as well as the assembly history itself of a galactic system.

\section*{Acknowledgments}
{ We thank the anonymous referee for his/her helpful comments.
Observations of M31 were carried out} with 
the Gran Telescopio CANARIAS (GTC), proposal ID: 
GTC39-17A (P.I. A.~Vazdekis). The authors thank dr. S.~Geier for the kind help he provided for the preparation of the phase2 material, and to perform the observations of M31.
FLB acknowledges the Instituto de Astrof\'isica de Canarias for the
kind hospitality when this project started.    FLB and AV acknowledge support from grant PID2019-107427GB-C32 from Ministerio de Ciencia e Innovaci\'on. FLB and AP acknowledge financial support from the INAF PRIN 1.05.01.85.11. 

%%%%%%%%%%%%%%%%%%%%%%%%%%%%%%%%%%%%%%%%%%%%%%%%%%
\section*{Data Availability}
The E-MILES SSP models are publicly available at the MILES website
(\url{http://miles.iac.es}). The Na-MILES models are available at the
same website (under the section "Other predictions/data").  The
CvD12a SSP models are available upon request to the
authors (see \url{https://scholar.harvard.edu/cconroy/projects}).
Data for the LB19 sample (programmes 092.B-0378, 094.B-0747,
097.B-0229; PI: FLB) can be downloaded from the ESO archive
(\url{http://archive.eso.org}).  Raw spectroscopic data for the bulge of M31
(programme GTC39-17A) are available from the GTC Public Archive
(\url{http://gtc.sdc.cab.inta-csic.es/gtc/jsp/searchform.jsp}).  The reduced radially-binned
spectra of M31 are available upon request to FLB.

%%%%%%%%%%%%%%%%%%%%%%%%%%%%%%%%%%%%%%%%%%%%%%%%%%



%\bibliographystyle{mnras}


%%%%%%%%%%%%%%%%%%%%%%%%%%%%%%%%%%%%%%%%%%%%%%%%%%

%%%%%%%%%%%%%%%%% APPENDICES %%%%%%%%%%%%%%%%%%%%%

\appendix

\section{Emission corrections}
\label{app:ecorr}

Absorption indices analyzed in the present work are corrected for contamination by (gas) emission lines. Besides the Balmer lines, \hb\ and $H\gamma$, contamination affects also the Fe5015 and Mgb absorptions, because of emission  from the [OIII]$\lambda\lambda4959,5007$ and [NI]$\lambda\lambda5198,5200$ doublets, respectively (see e.g.~\citealt{Sarzi:2006} and references therein). { Note} that, in the present work, we correct for the effect of emission on absorption features, while we postpone a detailed analysis of emission lines in the M31 OSIRIS spectra to a forthcoming paper. The correction is performed as follows:
\begin{description}
 \item[(i)] \hbo\ and \hgf\ are corrected with the same approach as in LB13 and LB16. For each radially binned spectrum, and each Balmer line, we perform spectral fitting on a 200~\AA-wide region around the line, with a 2SSP model~\footnote{The fitting allows for a variation of the ages and metallicities, as well as the relative light-fraction, of the two SSPs.}, by excluding the trough of the line, and fitting a Gaussian function to model the emission in the residual spectrum. The Gaussian fit is removed from the input spectrum to estimate the emission-corrected line-strength. Uncertainties on emission correction are estimated by taking into account noise in the input spectrum, by varying the degree of the polynomial modeling the continuum in the spectral fitting, as well as by varying the adopted IMF (i.e. the \gammab) of the 2SSP models.  The corrections turned to be significant (at more than 2.5~sigma level) for all radial bins, amounting, on average, to $\sim 0.3$~\AA\ and $\sim 0.1$~\AA\ for \hbo\  and \hgf, respectively.
 \item[(ii)] To correct Fe5015 and \mgb,  we performed spectral fitting with pPXF over the spectral region from $\lambda \sim 4800$ to $\sim 5300$~\AA, by excluding regions potentially affected by emission lines, i.e. regions of $\pm$100~\kms\ around the \hb$\lambda4861$ line, and around each line of the [OIII]$\lambda\lambda4959,5007$ and [NI]$\lambda\lambda5198,5200$ doublets. We modeled emission lines in the residual spectra with Gaussian functions, and removed the Gaussian profiles from the input spectra, to measure the emission-corrected line-strengths.
 On average, the corrections amount to $\sim -0.1$~\AA, and $+\sim 0.3$~\AA, for \mgb\ and Fe5015, respectively. { Note} that this method provides also an emission correction to \hbo\, that is similar ($\sim 0.05$~\AA\ lower) to that obtained from method (i).  However, to be consistent with our previous works (e.g. LB19), we decided to use corrections on \hbo\ and \hgf, from individual fits around each line, as described above (method i). Our values of the correction for emission contamination are roughly consistent with those of $0.29$~\AA, $-0.07$~\AA\ and $0.4$~\AA, reported by S10 for \hb\ (rather than \hbo), \mgb, and Fe5015, respectively.
\end{description}

\section{Correction of \nad\ for ISM absorption}
\label{app:ism}

\begin{figure}
 \begin{center}
\leavevmode
    \includegraphics[width=8cm]{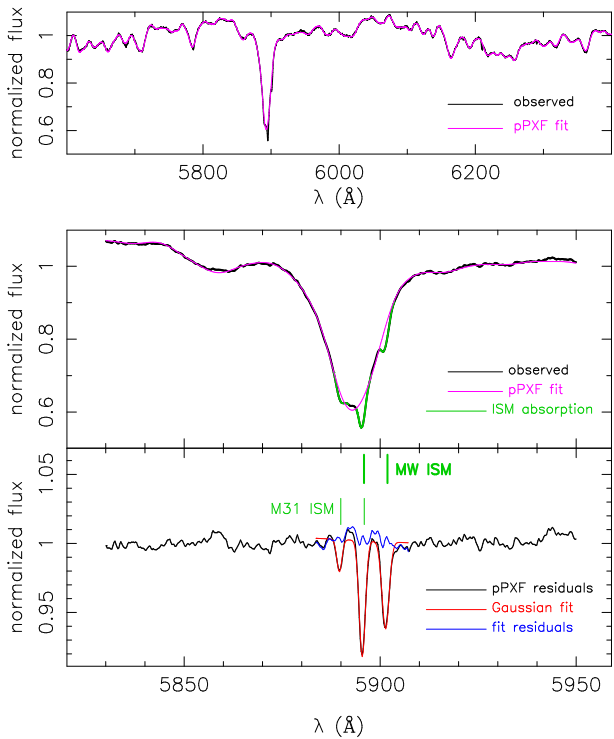}
 \end{center}
    \caption{Example of the procedure to correct NaD lines for ISM absorption, for the innermost spectrum of M31.
    Top panel: observed (black) and pPXF best-fit (magenta) spectrum, in the wavelength range from $\sim 5600$ to $\sim 6400$~\AA .
    Medium panel: the same as in the top panel but zooming into the spectral region around the NaD. The green curves mark regions potentially affected by ISM absorption (see below), that are masked out when running pPXF.
    Bottom panel: the black curve plots the observed spectrum normalized to the pPXF best-fit  around the NaD lines. { Note} the three absorptions at $\lambda \sim 5890$, $5895$, and $5902$~\AA, respectively. The thin and thick vertical segments, labeled as ``M31 ISM'' and ``MW ISM'', mark the expected positions of the NaD doublet (D2 and D1 lines) in the restframe of M31, and at the heliocentric velocity of M31 (about $-300$~\kms ), respectively. Regions potentially affected by ISM absorption (middle panel) are defined by windows $\pm100$~\kms--wide around the positions of the green segments. Red and blue curves show the multi-Gaussian best-fit to the ISM absorptions and the normalized residuals, respectively.  
     }
    \label{fig:NaDISM}
\end{figure}

The NaD stellar absorption is potentially affected by absorption from inter-stellar matter (ISM), both in the source restframe, as well as, for nearby systems, in our Galaxy. Although contamination can be severe for late-type galaxies~\citep{Jeong:2013}, none of the previous studies  has estimated the effect for the bulge of M31 (e.g.~\citealt{Z:15}). Thanks to the high quality of our OSIRIS spectroscopy, we are able to see the signature of ISM absorption on the profile of the NaD lines of M31. Fig.~\ref{fig:NaDISM} illustrates, for the innermost spectrum of M31,  the procedure we adopt to correct the equivalent width of NaD for ISM absorption. The observed spectrum of M31 is fitted with pPXF, over a region  $\sim 800$~\AA--wide around the NaD, as shown by the black and magenta curves in the top panel of the Figure. The middle and bottom panels show a zoom-in of the NaD absorption (middle), and the profile obtained by normalizing the NaD absorption with the pPXF best-fit solution (bottom). Three narrow ``residual absorptions'' are detected at $\lambda \sim 5890$, $5895$, and $5902$~\AA, roughly corresponding to the positions of the D2 and D1 lines ($\sim 5890$ and $5895$~\AA, respectively)  in the restframe of M31 (see the vertical thin green segments in the bottom panel), and at the heliocentric velocity of M31 (about $- 300$~\kms; see the thick green segments in the bottom panel). { Note} that in order to minimize the effect of ISM absorption on the best-fit of the stellar component, we run pPXF by masking out regions $\pm 100$~\kms--wide around the three residual absorptions shown in the middle panel of Fig.~\ref{fig:NaDISM} (see green lines in the middle panel).
We fit the residual absorptions with three Gaussian functions, and correct the input observed spectrum by dividing it with the best-fit Gaussians. The difference of NaD line-strength between the corrected and input spectra give the NaD correction for ISM absorption, $\rm \delta(NaD)$. 

\begin{figure}
 \begin{center}
\leavevmode
    \includegraphics[width=8cm]{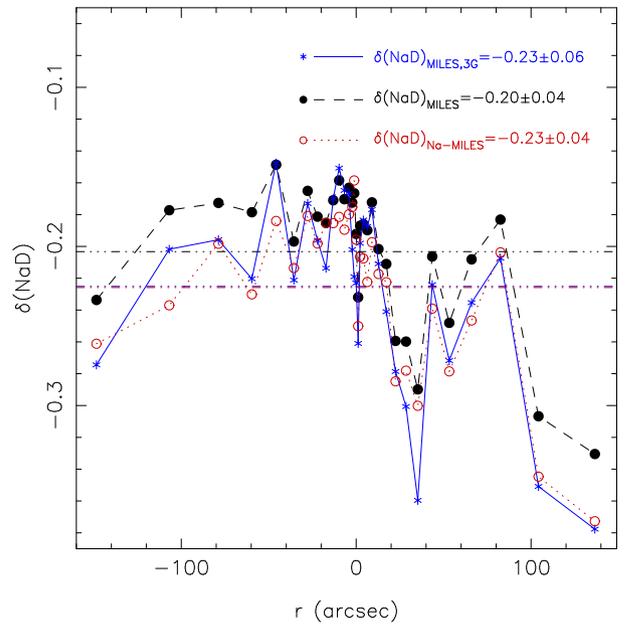}
 \end{center}
    \caption{Correction to the NaD line-strength for ISM absorption, $\delta(NaD)$, as a function of galacto-centric distance, $\rm R$, in units of arcsec. Different symbols and lines correspond to different methods to compute the correction, as labeled in the upper-right (see the text for details). Horizontal dot-dashed lines mark the median values of $\delta(NaD)$ from different methods.
     }
    \label{fig:NaDcorr}
\end{figure}

The value of $\rm \delta(NaD)$ is negative, as expected by the fact that ISM contamination makes the NaD stronger with respect to that of the pure stellar component. Fig.~\ref{fig:NaDcorr} plots $\rm \delta(NaD)$ as a function of radial distance to the center of M31. To explore possible systematics in the correction, we repeated the entire procedure by running pPXF with two different sets of templates, i.e. MILES stars as well as Na-EMILES SSP models. Also, since the ISM absorption at $\lambda \sim 5890$~\AA\ is much weaker than the others, and is barely seen in some spectra (especially for the outer, lower S/N, radial bins), we repeated the procedure by fitting either three or two Gaussian functions (in the latter case only  the residual absorptions at $\lambda \sim 5895$ and $5902$~\AA\ were fitted). Different methods are shown with different colours, symbols, and line types in Fig.~\ref{fig:NaDcorr}, with average values from different methods (see the labels) marked with horizontal lines. For all methods, the correction is significant, and amounts to about $-0.2$~\AA. { Note} that this value is negligible with respect to the \nad\ radial gradient as measured in  the present work (see panel g of Fig.~\ref{fig:indices}). As a conservative approach
we averaged out all results shown in Fig.~\ref{fig:NaDcorr}, and applied a constant correction of $-0.215 \pm 0.06$\AA\ to all NaD line-strengths analyzed in the present work~\footnote{In general, the correction procedure is hampered by the fact that NaD airglow lines do also overlap to the MW ISM absorption lines. This contribution is negligible in the innermost region, where the signal from the galaxy is prominent, while it becomes more important in the outermost bins, hampering a precise estimate of the radial variation of $\rm \delta(NaD)$.  }. 
%We notice that neglecting the effect of $\delta(NaD)$ would lead %to over-estimate the abundance of \nafe\ (by $\sim $~dex, hence  

\section{Age, \zh, and \mgfe\ grids}
\label{app:azagrids}

\begin{figure}
 \begin{center}
\leavevmode
    \includegraphics[width=8cm]{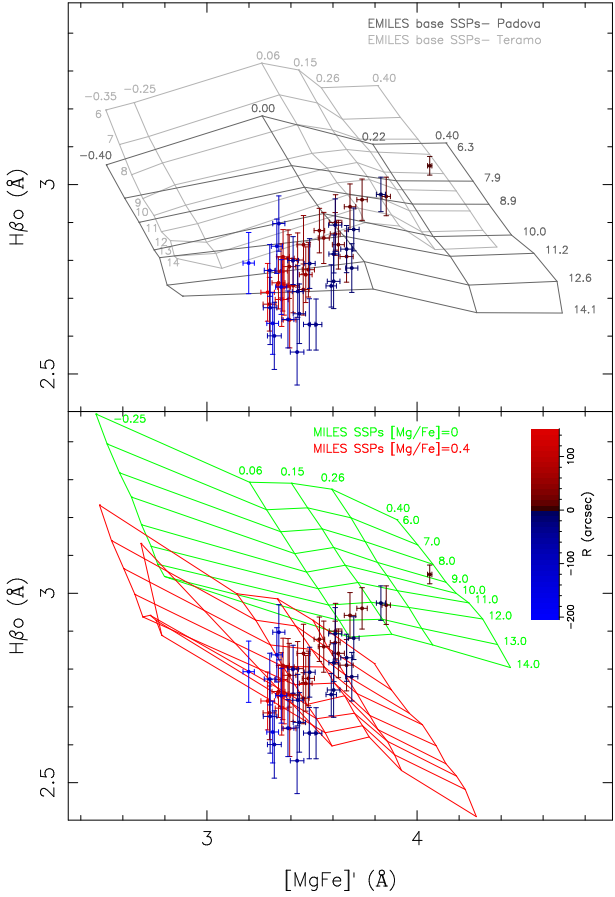}
 \end{center}
    \caption{The age indicator \hbo\ is plotted against the metallicity indicator \mgfep, for EMILES base (top) and $\alpha$--MILES (bottom) SSP model predictions. In both panels, observed line-strengths for the M31 bulge are shown with dots and error bars (1-sigma confidence levels), with colours varying from blue (negative values of the radial distance $R$) through red (positive values of $R$), as shown in the bottom-panel inset bar. In each panel, the grids show SSP model predictions with  varying age and metallicity, along the horizontal and vertical directions, respectively, as shown by the labels. In the top panel, dark- and light-grey grids correspond to Padova and Teramo models. For $\alpha$--MILES models (bottom), labels are shown only for the scaled--solar (green) models. { Note} that the grids correspond to different values of age and metallicity, depending on the models' isochrones (Teramo vs. Padova). The age range from $\sim 6$ to $\sim 14$~Gyr, while metallicity varies from about $-0.25$~dex ($-0.4$) for Teramo  (Padova) models, up to $+0.4$~dex. Predictions for the highest metallicity values are obtained by extrapolation. 
    }
    \label{fig:M31_HboMgFep}
\end{figure}

Fig.~\ref{fig:M31_HboMgFep} illustrates the estimate of age and metallicity for the M31 spectra through the \hbo--\mgfep\ diagram, using different SSP models, i.e. base EMILES models with BaSTI and Padova isochrones~\footnote{{ Note} that we refer to SSP models based on BaSTI isochrones as ``Teramo'' models.} (dark- and light-grey grids in the top panel), and $\alpha$--MILES scaled-solar (green) and $\alpha$-enhanced (red) models in the bottom panel. Since around solar metallicity, the BaSTI isochrones are hotter than the Padova ones (see~\citealt{Pietrinferni04}; V15), Teramo base models are shifted towards higher \hbo\ values than Padova (see the top panel of the Figure). Hence, Teramo models tend to provide older ages (by $\sim 1$--$2$~Gyr) than Padova models, as shown in Fig.~\ref{fig:agemetalfa}. However, when using $\alpha$--MILES models (see bottom panel of the Figure), the effect of the isochrones is counteracted by the fact that \hbo\ is expected to decrease with \mgfe\ (see the red and green grids in the bottom panel). Since the M31 spectra are $\alpha$-enhanced, ages obtained with $\alpha$--MILES models do not differ significantly from those obtained with Padova base models, as shown in Fig.~\ref{fig:agemetalfa}. We { note} that 
the metallicities obtained with Teramo (base) models are also slightly higher than those for Padova, as the two sets of models differ for the reference solar metallicity (see V15). 

\begin{figure}
 \begin{center}
\leavevmode
    \includegraphics[width=8cm]{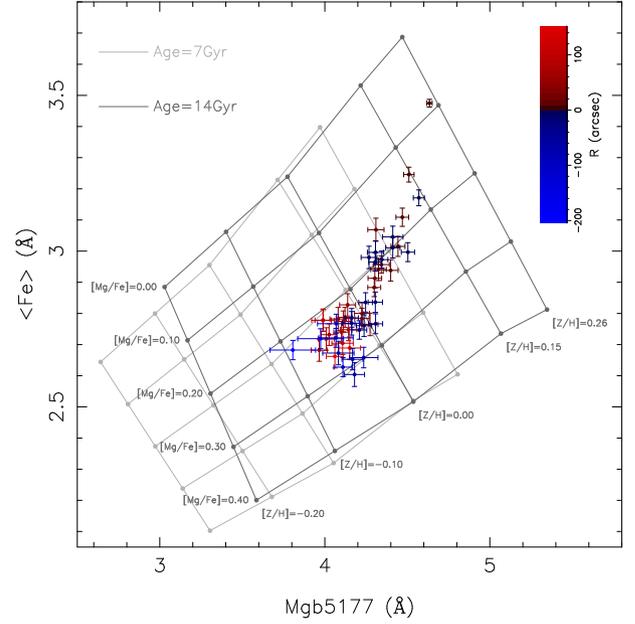}
 \end{center}
    \caption{The \fem\ spectral index is plotted as a function of \mgb. Observed line-strengths for the M31 bulge are shown with coloured dots, as in Fig.~\ref{fig:M31_HboMgFep}. The dark- and light- grey grids show SSP model predictions of $\alpha$--MILES models, for an age of $7$ and $14$~Gyr, respectively, corresponding to the innermost and outer radial bins of M31.
    The grids show the effect of varying metallicity, from $-0.2$ to $+0.26$~dex, and \mgfe , from 0 to $0.4$~dex, as labeled.
    { Note} that points for \mgfe$=0.1$, $0.2$, and $0.3$~dex in the grids, are obtained by linear interpolation of the models. 
     }
    \label{fig:M31_FemMgb}
\end{figure}

For what concerns the estimates of \mgfe\ (bottom panel of Fig.~\ref{fig:agemetalfa}), we find a decrease of \mgfe\ in the center of the bulge, which is more pronounced for $\alpha$--MILES models, and barely seen with base models (where we use the ``proxy'' for \mgfe; see LB13 and V15). Fig.~\ref{fig:M31_FemMgb} plots the \fem\ index as a function of \mgb. In this diagram (see e.g.~\citealt{TMB:03}), the effect of metallicity and age are almost parallel, pushing the line-strength of both indices to higher values. On the other hand, increasing \mgfe\ tends to increase \mgb\ and decrease \fem, so that the effect of \mgfe\ is orthogonal to that of age and metallicity. As seen in the Figure,  the radial gradient for the M31 data-points is along a direction of decreasing metallicity and increasing \mgfe, with \mgfe\ values between $0.1$~dex for the innermost radial bin, and $\sim 0.25$~dex for the outermost bins ($r \gtrsim 100$''). Hence, Fig.~\ref{fig:M31_FemMgb} confirms the presence of a (mild) positive radial gradient of \mgfe\ for the bulge of M31. { Note} that a lower \mgfe\ in the center of M31 has also been reported by~\citet{Davidge:1997}.

\section{The \tioi\ and \tioii\ mismatch at $\rm R=0$}
\label{app:tios}

In the innermost radial bins, we are able to match all observed indices of M31, but \tioi\ and \tioii\  (see Fig.~\ref{fig:indices2} and Sec.~\ref{sec:bestfit}). The mismatch stems from the fact that EMILES models predict both \tioi\ and \tioii\ to increase with age and IMF slope, while being independent of \zh\ (for \zh$\gtrsim 0$). On the other hand, the observed TiO's show a sharp rise towards the center of the bulge (within $\sim 40$''), where age is younger, \gammab\ higher, and metallicity is above solar (see Fig.~\ref{fig:agemetalfa} and~\ref{fig:gammab}). Therefore, the best-fit values of both TiO's underestimate the observed line-strengths at $\rm R \sim 0$. { Note} that the mismatch is unlikely due to the effect of non-solar abundance ratios. In fact, the discrepancy between observed and model indices does not disappear when including individual abundance ratios in the fitting procedure (methods I and K). Moreover, TiO indices are expected to increase with \afe, while \mgfe\ abundance is found to decrease in the bulge center (see Fig.~\ref{fig:agemetalfa}). 

In the center of M31, stellar metallicity is significantly above solar, with values as high as \zh$=+0.4$~dex. For \zh\ above solar, EMILES models are actually computed for \zh$=+0.15$, $+0.26$, and $+0.4$, respectively. However, predictions are safe up to \zh$=+0.26$, while they become unsafe at \zh$=+0.4$, due to the dearth of high metallicity stars in the input stellar  libraries (see V15). For this reason, in our reference approach (e.g. LB17 and LB19), we do not use models with \zh$+0.4$, but rather perform a linear extrapolation of the models above $+0.26$ (see Sec.~\ref{sec:spmodels}). 
To investigate if this extrapolation might be responsible for the TiO mismatch in the center of M31, we created a set of 
EMILES experimental models, with the main aim of improving predictions at \zh$=+0.4$~dex. First, we { note} that, when computing an SSP model,  
the iron metallicity ([Fe/H]), total metallicity (\zh), and \mgfe\ of stars to be attached to the isochrone, are related by the equation:
\begin{equation}
  \rm [Z/H] = [Fe/H] + \beta \cdot [Mg/Fe],
\end{equation}
where the constant $ \beta$ is $ \sim 0.75$ for $\alpha$-MILES models (see V15). Therefore, to compute  a given SSP, one needs  lower  [Fe/H] to achieve the required \zh,  if \mgfe$>0$, meaning that one can compute better quality models if both \zh\ and \mgfe\ are above solar. Unfortunately, most of MILES stars at [Fe/H]$\gtrsim 0$ have \mgfe\ around solar, meaning that, in principle, if one wants to create an $\alpha$-enhanced model at $\zh \gtrsim 0$, one has to rely on theoretical differential corrections, with all the uncertainties inherited from theoretical models of stellar atmospheres (see V15). A way out is that of using {\it the 
scatter} in \mgfe\ of stars at given [Fe/H], to create, by interpolation, empirical stellar spectra with \mgfe$>0$. Therefore, if we do not ``travel too much above'' \mgfe$\sim 0$, we can construct better-quality $\alpha$-enhanced models at supersolar metallicity, entirely based on empirical stellar spectra. We decided to follow this approach, by computing a set of EMILES experimental {\it empirical} models with \mgfe$=+0.1$~\footnote{These $\alpha$-enhanced models are computed assuming scaled-solar isochrones, as $\alpha$-enhanced Teramo isochrones are available only for \mgfe$=0.4$. However, as discussed in V15,  the effect of $\alpha$-enhanced isochrones is secondary compared to that of \mgfe\ for stellar spectra.}. 

\begin{figure}
 \begin{center}
\leavevmode
    \includegraphics[width=8cm]{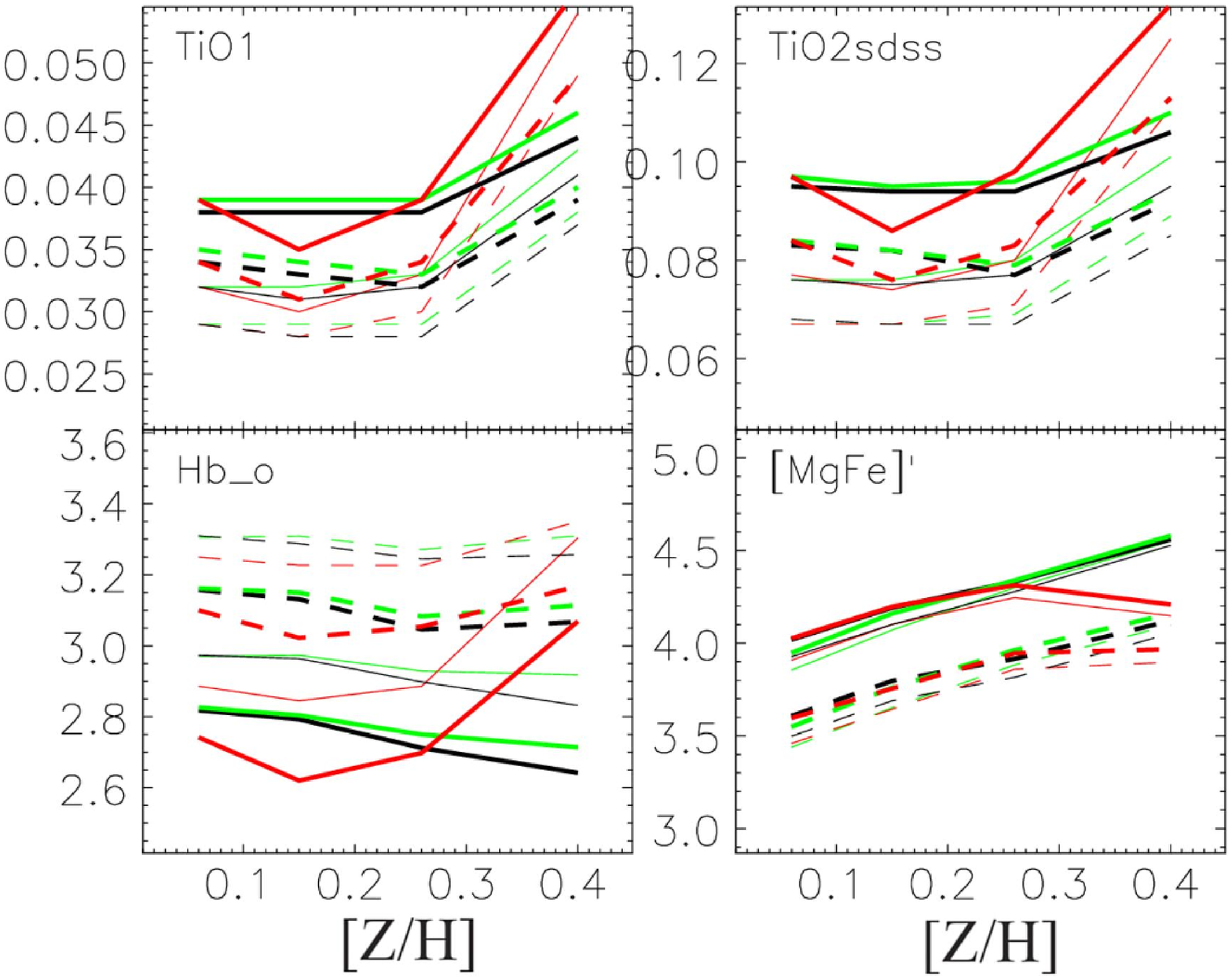}
 \end{center}
    \caption{SSP model predictions for \tioi\ (panel a), \tioii\ (panel b), \hbo\ (panel c), and \mgfep\ (panel d), as a function of total metallicity \zh . Dashed and solid lines refer to models with an age of 7 and 14~Gyr, typical for the innermost and outermost bins of M31, respectively. Thin and thick lines correspond to a Kroupa-like (\gammab$=1.3$), and a bottom-heavy (\gammab$=3$) IMF, respectively, while black, green, and red curves plot predictions for base, scaled-solar, and $\alpha$-enhanced EMILES testing models, with \mgfe$=0.1$, specifically computed for this project (see the text). The safe range of EMILES is up to \zh$_{\rm safe}=+0.26$ (see V15), with the \mgfe$=0.1$ testing models having better quality above this metallicity threshold. { Note} the trend of \tioi\ and \tioii\ to be constant up to \zh$_{\rm safe}$, and then increase significantly with \zh\ for \zh$>$\zh$_{\rm safe}$.
     }
    \label{fig:TiO_MH}
\end{figure}

In Fig.~\ref{fig:TiO_MH} we plot SSP model predictions for both \tioi\ and \tioii , as well as for the age-sensitive index, \hbo, and for the total metallicity indicator \mgfep, as a function of \zh. We consider only models computed with Teramo isochrones. Black and green curves in the Figure show EMILES base and scaled-solar models, respectively, while red curves correspond to the experimental empirical models with \mgfe$=+0.1$. 
Dashed and solid lines in the Figure refer to models with age of 7 and 14~Gyr, respectively, typical for the innermost and outermost bins of M31. Thin and thick lines are for a Kroupa-like (\gammab$=1.3$), and a bottom-heavy (\gammab$=3$) IMF, respectively,
Fig.~\ref{fig:TiO_MH} shows that both TiO indices do not depend significantly on metallicity from $\rm [Z/H] \! \sim \! 0$ to $\rm [Z/H] \! \sim \! +0.26$. Therefore, when extrapolating the models above \zh$=+0.26$, as in our reference approach, one predicts both TiO's to be constant all the way up to \zh$=+0.4$~dex, and above.
On the contrary, base and scaled-solar models suggest an increase of  0.01--0.015~mag (depending on age and IMF slope) for \tioii, and about 0.005--0.01~mag for \tioi, from \zh$=+0.26$ to \zh$=+0.4$.  Although models at \zh$=+0.4$ are unsafe, this trend suggests that our linear extrapolation might be underestimating the TiO line-strengths at \zh$=+0.4$, explaining the mismatch in the center of  M31. The experimental empirical models with \mgfe$=+0.1$ seem to confirm that this is actually the case, predicting an even stronger increase of both TiO's with \zh\ (see the red curves in the Figure for \zh$>0.26$), with respect to scaled-solar and base models. However, at the same time, the models with \mgfe$=+0.1$ predict an increase (decrease) of \hbo\ (\mgfep) with \zh\ (for \zh$>0.26$), that contrasts with the expectation of \hbo\ (\mgfep) being insensitive to metallicity (\mgfe), hence casting some doubts about the reliability of predictions in Fig.~\ref{fig:TiO_MH}. Therefore, while the present analysis seems to support the conclusion that the mismatch of TiO observed and model line-strengths in the center of M31 is due to an increase of both \tioi\ and \tioii\ with metallicity, in the highly supersolar regime, current stellar population models do not allow us to draw firm conclusions.

In order to account for a possible increase of \tioi\ and \tioii\ at \zh$>0.26$, we modified our best-fitting method I (see Sec.~\ref{sec:IMFfitting}). For \zh$>0.26$, we assume a linear parametrization of TiO1 and TiO2 as a function of \zh,  with slopes of 
 $\rm \Delta_Z (TiO1)= \delta (TiO1)/\delta([Z/H])$ and $\rm \Delta_Z (TiO2)= \delta (TiO2)/\delta([Z/H])$, respectively. 
The values of  $\rm \Delta_Z (TiO1)$ and $\rm \Delta_Z (TiO2)$ are assumed to be independent of age and IMF slope, and are treated as two extra fitting parameters.
From the data of M31 at $\rm R=0$, we obtain $\rm \Delta_Z (TiO1)=0.07\pm0.014$~{ mag/dex} and $\rm \Delta_Z (TiO2)=0.23\pm0.014$~{ mag/dex}. { Note} that these values are remarkably consistent with those one can estimate from the experimental
models in Fig.~\ref{fig:TiO_MH}, i.e. $\rm \Delta_Z (TiO1) \sim 0.1$~{ mag/dex} and $\rm \Delta_Z (TiO2) \sim 0.25$~{ mag/dex}, respectively. Also, the best-fitting line-strengths for the modified method I, plotted as green asterisks in Figs.~\ref{fig:indices} and~\ref{fig:indices2}, match all indices of M31 at $\rm R=0$, including the two TiO's~\footnote{As a further test, we also included the NIR TiO index $\rm TiO0.89$ in the fitting (see LB17), finding no significant differences in the estimated values of $\rm \Delta_Z (TiO1)$ and $\rm \Delta_Z (TiO2)$, as well as in the best-fit value of \gammab. }. More important for the purpose of the present paper,  when accounting for 
$\rm \Delta_Z (TiO1)$ and $\rm \Delta_Z (TiO2)$, the best-fitting IMF slope, \gammab, changes only slightly with respect to method I, decreasing from $\sim 2.4$ to $\sim 2.3$. We conclude that, while better models are certainly needed in the very-high metallicity regime, our results are robust against the model uncertainties at high \zh.

\section{Kinematics}
\label{app:kin}

\begin{figure}
 \begin{center}
\leavevmode
    \includegraphics[width=8cm]{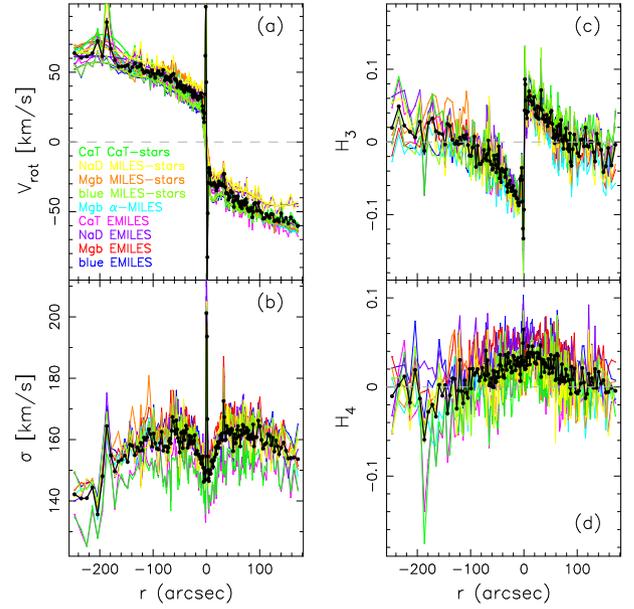}
 \end{center}
 \caption{The kinematics profiles for the bulge of M31, i.e. rotation velocity \VROT (panel a), velocity dispersion \SIG\ (panel b), as well as \HT\ and \HF\ (panels c and d), are plotted as a function of galactocentric distance, $\rm R$, in units of arcsec. Lines with different colours correspond to results obtained by running pPXF on different spectral ranges, i.e. the ``blue'', ``Mgb'', ``NaD'', and ``CaT'' regions (see the text), and with different sets of templates, i.e. EMILES SSP models, MILES and CaT stars, and $\alpha$--MILES SSP models, as labeled in the lower--left of panel a. { Note} that the labels report the fitted spectral range, followed by the adopted set of templates. 
 { Note} the small offset of $\sigma$ (about $-10$~\kms), for results corresponding to the CaT spectral region (see text).
 Results from different methods are median-combined into final kinematics profiles, shown as black curves.
 }
    \label{fig:M31_KIN_METHODS}
\end{figure}

To extract the kinematics of the M31 bulge, we have performed an adaptive binning of the OSIRIS spectra as a function of radial distance, assuming a minimum $S/N$ ratio of 25~\AA$^{-1}$ (see Sec.~\ref{subsec:binning} for details). We verified that adopting a minimum $\rm S/N$ of 15, rather than 25, does not change significantly our results.
We measured  rotation velocity \VROT , velocity dispersion, \SIG , and the higher moments of the line-of-sight (LOS) velocity distribution, \HT\ and \HF ,
by running the software {\sc pPXF}~\citep{Cap:2004, Capp:17}. We ran pPXF within four different spectral ranges, i.e. the ``blue'' ($\lambda\lambda3900$-$4900$~\AA), ``Mgb'' ($\lambda\lambda4750$-$5750$~\AA), ``NaD'' ($\lambda\lambda5700$-$6200$~\AA), and ``CaT'' ($\lambda\lambda7670$-$8850$~\AA) spectral regions, adopting two sets of templates, i.e. EMILES SSP models with varying age and metallicity (see Sec.~\ref{sec:spmodels}), as well as stellar spectra from the MILES~\citep{MILESI} (for blue, Mgb, and NaD ranges) and CaT~\citep{CATI} (for the CaT range) libraries. For the Mgb range, we also ran pPXF with $\alpha$--MILES models (including both scaled-solar and $\alpha$--enhanced models). Fig.~\ref{fig:M31_KIN_METHODS} compares the kinematics radial profiles obtained for different methods. Indeed, we find an excellent agreement among different spectral ranges, and different templates, but for a small offset in $\sigma$ (amounting to about $-10$~\kms) for the CaT spectral range (see pink and light-green curves in panel b of the Figure). Since we obtained consistent results among methods, we median-combined different results into final kinematics profiles, shown as black curves in Fig.~\ref{fig:M31_KIN_METHODS}.

\begin{figure}
 \begin{center}
\leavevmode
    \includegraphics[width=8cm]{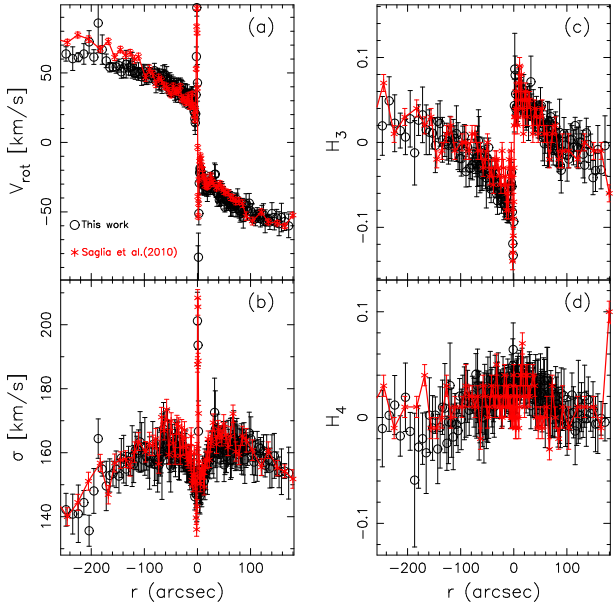}
 \end{center}
 \caption{
   Comparison of our kinematics for the bulge of M31 (black circles and error bars, as in Fig.~\ref{fig:M31_KIN_METHODS}), with that obtained by \citealt{Saglia:2010}, shown as red stars. Panels a, b, c, and d correspond to the same kinematics profiles as in Fig.~\ref{fig:M31_KIN_METHODS}. { Note} the excellent agreement between the two sets of measurements.
        }
    \label{fig:M31_KIN}
\end{figure}

Fig.~\ref{fig:M31_KIN} shows that our kinematics for the M31 bulge (black curves) is in excellent agreement with that obtained by~S10. Rotation is detected at all radii probed in the present work, and  correlates with \HT . We confirm that within a region of $\sim 100$'', the galaxy has a velocity dispersion of $\sim 160$~\kms, dropping down (by $\sim 10$~\kms ) towards the center (see \citealt{Opitsch:2018} and references therein). On the other hand, in the very central bins, { the velocity dispersion} increases steeply -- a signature of the central supermassive black hole.
Within $\sim 100$'', the \HF\ is slightly positive, with a typical value of $\sim 0.02$, again in very good agreement with S10.
{ Note} that in our data, we also see a slight trend of \HF\ decreasing outwards, which is not clearly seen in S10 data.

\begin{figure}
 \begin{center}
    \leavevmode
    \includegraphics[width=8cm]{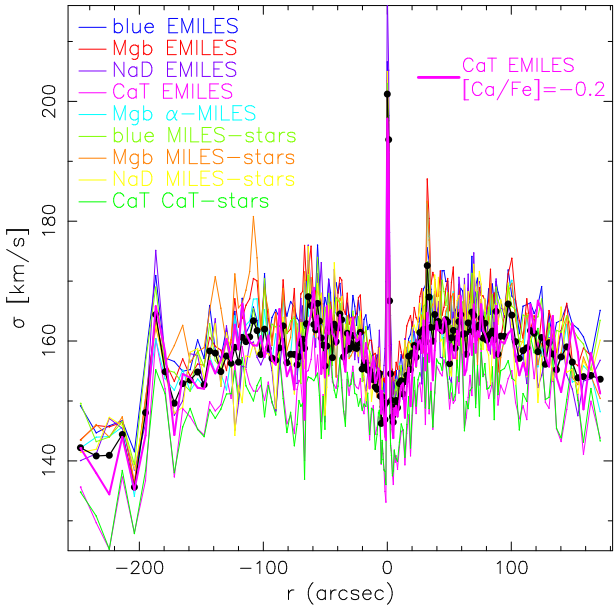}
 \end{center}
 \caption{The same velocity dispersion profiles as in panel c of Fig.~\ref{fig:M31_KIN_METHODS} (see labels in the upper--left), are compared to the \SIG\ profile obtained with EMILES models, modified for an abundance pattern of \cafe=$-0.2$~dex, plotted as a thick pink curve (see the text). { Note} that for models with \cafe$=-0.2$~dex, we match the kinematics obtained in the bluer spectral ranges, removing the $\sigma$ offset seen in panel b of Fig.~\ref{fig:M31_KIN_METHODS}. 
 }
    \label{fig:M31_SIGMA_CAFE}
\end{figure}

Since the observed line-strengths for CaT are below the predictions of EMILES models (see, e.g., the red curve in panel f of Fig.~\ref{fig:indices}), the small offset in \SIG\ for the CaT spectral range (see above) might be due to template mismatch. In order to further explore this point, we re-ran pPXF in the CaT region with EMILES models, applying a response for an abundance pattern of \cafe$=-0.2$ (as estimated from CvD12a models). In this case, results for the CaT are in excellent agreement with those obtained for the other spectral ranges, as shown by the thick magenta line in Fig.~\ref{fig:M31_SIGMA_CAFE}. We conclude that template mismatch is the most likely explanation for the (small) offset of \SIG\ in the CaT region.

%%%%%%%%%%%%%%%%%%%%%%%%%%%%%%%%%%%%%%%%%%%%%%%%%%

% Don't change these lines
\bsp	% typesetting comment
\label{lastpage}
\end{document}